\documentclass[acmsmall, screen]{acmart}
\AtBeginDocument{%
  }

\setcopyright{none}
\copyrightyear{2026}
\acmYear{2026}
\acmDOI{} 

\acmISBN{} 

\usepackage{graphicx} 
\usepackage{amsmath,amssymb,amsthm}
\usepackage{mathtools}
\usepackage{url}
\usepackage{tikz}
\usetikzlibrary{arrows.meta, patterns}
\usetikzlibrary{positioning, arrows.meta, fit, backgrounds}
\usepackage{xcolor}
\usepackage{booktabs}
\usepackage{pgfplots}
\pgfplotsset{compat=1.18}
\usepackage{pgfplotstable}
\usepackage{doi}

\usepackage{multirow}
\usepackage{mdframed}
\usepackage{hyperref}
\hypersetup{hidelinks, pdfborder={0 0 0}}
\usepackage{enumitem}
\usepackage{algorithm}
\usepackage{algpseudocode}

\newtheorem{observation}{Observation}
\usepackage{comment}

\usepackage{cite}

\begin{document}

\title{Structure-Preserving Error-Correcting Codes for Polynomial Frames}

\author{Baigang Chen}
\affiliation{%
  \institution{University of Washington}
  \country{USA}}
\email{chen9464@umn.edu}

\author{Dongfang Zhao}
\affiliation{%
  \institution{University of Washington}
  \country{USA}}
\email{dzhao@cs.washington.edu}

\begin{abstract}
Modern FFT/NTT analytics, coded computation, and privacy-preserving ML interface routinely move polynomial frames across NICs, storage, and accelerators. However, even rare silent data corruption (SDC) can flip a few ring coefficients and cascade through downstream arithmetic. Conventional defenses are ill-matched to current low-latency pipelines: detect-and-retransmit adds RTTs, while byte-stream ECC ignores the algebraic structure and forces format conversions. To that end, we propose a structure-preserving reliability layer that operates in the encoded data's original polynomial ring, adds a small amount of systematic redundancy, and corrects symbol errors/flagged erasures without round-trip or format changes. We construct two complementary schemes: one for odd length $N_{odd}$ via a Hensel-lifted BCH ideal with an idempotent encoder, and one for power-of-two length $N_{2^m}$ via a repeated-root negacyclic code with derivative-style decoding. In particular, to stay robust against clustered errors, a ring automorphism provides in-place interleaving to disperse bursts. Implementation wise, on four frame sizes $N\!=\!1024, 2048, 4096, 8192$, we meet a per-frame failure target of $10^{-9}$ at symbol error rates $10^{-6}\text{--}10^{-5}$ with $t\!=\!8\text{--}9$, incurring only $0.20\%\text{--}1.56\%$ overhead and tolerating $\sim\!32\text{--}72$\,B unknown-error bursts (roughly doubled when flagged as erasures) after interleaving. By aligning error correction with ring semantics, we take a practical step toward deployable robustness for polynomial-frame computations from an algebraic coding perspective.

\end{abstract}

\keywords{}

\received{}
\received[revised]{}
\received[accepted]{}

\maketitle
\section{Introduction}

Modern data systems often move polynomially encoded data between computing stages, for example, FFT/NTT-centric analytics~\citep{BradburyDruckerHillenbrand21} and accelerators in privacy-preserving ML~\citep{GuardNN-DAC22}, such as CKKS for approximate arithmetic~\citep{CKKS17}, PPML frameworks like CryptoNets~\citep{CryptoNets16} and Gazelle~\citep{Gazelle18}. Further, there exists coded computation that explicitly encodes tasks via polynomials such as Polynomial Codes~\citep{PolynomialCodes17}. In these settings, each data frame is a fixed-length array of coefficients stored in a structured polynomial ring over integers modulo a power of a prime~\citep{Banerjee2007PolynomialWSN}, commonly selected as two. As frames traverse network interface controllers (NICs)~\citep{BairavasundaramSIGMETRICS07LSE}, storage tiers~\citep{Khan2024CloudTier}, and accelerators, silent data corruption (SDC)~\citep{Dixit21SDC} can flip a few coefficients despite existing protections: large-scale DRAM field studies report non-negligible error rates~\citep{Schroeder2009DRAMWild}, production reports document CPU/logic-level SDCs that escape hardware reporting~\citep{Dixit21SDC}, and recent fleet-wide analyses characterize SDCs across over a million processors~\citep{WangSOSP23SDC}; on the storage path, latent sector errors remain a practical failure mode~\citep{BairavasundaramSIGMETRICS07LSE}. Even a handful of flipped symbols can derail later stages because polynomial pipelines rely on arithmetic among all coefficients—e.g., element-wise products in the transform domain must map back to valid ring products—so small perturbations propagate in non-obvious ways~\citep{GlitchFHE25}. 

Given the fact that SDCs appear with non-negative probability and have non-trivial damage to the reliability of data transportation, there are several techniques to defend against SDCs. However, common safeguards do not fit the workflow of current trends where the workspace is commonly computation-heavy heavy such as privacy-preserving machine learning pipelines~\citep{Xu2021PPMLSurvey} as well as FHEs including BGV~\citep{BGV2012ITCS} and TFHE~\citep{Chillotti2020TFHE}. For more general cases when the application requires multi-server computation or time sensitivity, those existing techniques almost fall apart. One of the most common strategies, hash and retransmit, detects corruption but forces a round trip and stalls the pipeline~\citep{RFC793, RFC9000, RFC9002, Zhang2010ZFSIntegrity}. Generic error correcting codes on raw bytes, such as BCH or Reed–Reed-Solomon~\citep {BoseChaudhuri1960, ReedSolomon1960}, protect the bytestream but ignore the algebra that later stages rely on. They require packing and unpacking steps, and break the desirable property that computation should preserve the structure of the encoded frame. Our goal is a reliability layer that rides with the frame and does not interrupt the pipeline. It should fit naturally into the existing workflow for polynomial frames, correct symbol errors and any positions flagged as erasures during transport or storage, and preserve the ring semantics so that linear and multiplicative stages do not destroy code membership.

To satisfy the above desirable properties, protection should operate inside the data-reside domain that real deployments already use and should be based on strong algebraic tools instead of protocol design. Aiming aforementioned goals, we add systematic redundancy in the polynomial domain: the encoder maps an input frame to a protected codeword and, when burstiness is expected, applies an automorphism-based interleaver that redistributes locality while preserving code membership. On receive, the decoder forms structure-aware  syndromes~\citep{Castagnoli1991RepeatedRoot} and performs bounded-distance errors/erasures correction up to a configured budget; the corrected frame is then forwarded to the next stage. Parameter selection ties the correction budget to the SDC model and per-frame failure target. Because the encoder respects structural arithmetic, all linear operations preserve membership; and, in one instantiation, multiplication also preserves membership, achieving compatibility for any circuits without re-encoding~\citep{Boemer2019nGraphHE2, CKKS2017}.

\paragraph{Contributions}
\begin{itemize}
  \item \textbf{A ring-compatible reliability component for polynomial-frame pipelines.}
  We design a drop-in, non-interactive component that corrects symbol errors and flagged erasures during transport/storage. Our component, presented as a codeword,  preserves the native ring semantics of the data. In particular, linear operators always preserve code membership; for odd~$N$, our lifted-BCH instantiation also preserves multiplicative stages, allowing the system component to follow data across compute boundaries without decoding and re-encoding.

  \item \textbf{Algorithms for common frame lengths.} We present two concrete encoding/decoding algorithms realized as ideals in a polynomial ring from theory foundations in algebraic coding theory: (i) a repeated-root negacyclic code for power-of-two lengths with derivative-style syndromes, and (ii) a Hensel-lifted BCH code with an idempotent encoder for odd lengths. For bursty faults, we add an automorphism-based interleaver that reindexes coefficients in place to disperse localized corruption, and such an interleaving construction is compatible with our codeword structure naturally.

  \item \textbf{Quantitative robustness under explicit fault models.}
  Under i.i.d.\ symbol flips with rates $p\!\in\![10^{-6},10^{-5}]$, parity budgets of $t\!=\!8$–$9$ achieve per-frame failure $\approx 10^{-9}$; across $N\!\in\!\{1024, 2048, 4096, 8192\}$, overhead is $0.20\%\,$–$\,1.56\%$. With the automorphism interleaver, the same parameters tolerate $\sim\!32$–$72$\,B unknown-error bursts (roughly doubling when positions are flagged as erasures via the standard $2\tau+\rho\le 2t$ rule).

\end{itemize}

\paragraph{Roadmap.}
Section~\S2 reviews some of the most related works in the literature. Section~\S3 fixes notation and briefly discusses the preliminaries.
Section~\S4 describes the system workflow and application programming interface (API). Section~\S5 details the two constructions: a repeated\textendash root negacyclic code for power\textendash of\textendash two lengths with derivative\textendash style syndromes, and a Hensel\textendash lifted BCH ideal for odd lengths with an idempotent encoder; it also introduces the in\textendash ring automorphism interleaver for burst dispersion.
Section~\S6 reports benchmarks regarding parameter sizing, overhead vs.\ frame length, encode/decode costs, and burst tolerance. Section~\S7 concludes and outlines future work.

\section{RELATED WORK}

This section contrasts our design with (i) error-correcting codes commonly used to protect data in practice, and (ii) systems-level integrity and recovery mechanisms that operate around the data path. Concretely, our reliability layer sits immediately above serialization, adds a small, fixed amount of symbol-level redundancy, optionally applies an in-place coefficient permutation, and removes both at the receiver or some middle "gas stop" depending on parameter choosing. Because purposed code families live in the same ring as the application data, linear stages inherently preserve protection. In the lifted-BCH case, multiplicative stages are preserved as well. This keeps the reliability-gaining process non-interactive and avoids format conversions, while still leveraging existing CRCs as erasure hints when present. In deployments, operators choose $(N,k,t)$ so that overhead (exactly $2t$ symbols per frame) and decoder cost match target SDC rates and data rates.

\subsection{Error-Correcting Codes}

\subsubsection{Byte-stream ECC} Reed–Solomon (RS) codes over bytes are the default in storage and wide area network (WAN) transport because they are maximum distance separable (MDS) and easy to deploy at object or chunk granularity. Production systems typically use $(k{+}r)$ configurations to balance storage overhead (30–50\%) against two–four symbol correction~\citep{Muralidhar2014f4,Sathiamoorthy2013Xorbas,Huang2012LRCSlides}. HDFS~\citep{HDFS-EC} and Ceph~\citep{Ceph-EC} expose these as erasure pools; cloud object stores split large objects into stripes with per-stripe RS parity. These deployments operate on raw byte blocks: the encoder reads $k$ contiguous chunks, generates $r$ parity chunks, and writes $(k{+}r)$ total. Applying RS to polynomial frames requires packing the ring elements into byte buffers and unpacking after recovery. That conversion breaks the property that a downstream linear or multiplicative stage preserves structure, so systems either accept the conversion cost or push protection to the edges. Low-density parity-check (LDPC)~\citep{BreuckmannEberhardt2021QLDPC} and related sparse-graph codes~\citep{Balitskiy2021LPNonlinear} dominate high-throughput links such as Wi-Fi, 5G NR, and DOCSIS~\citep{Fanari2021Coding80211ax}. They are decoded with belief propagation and tuned for fixed bit error rates and target SNRs. In practice, they protect PHY/MAC codewords~\citep{IEEE80211ax2021}, which are not application-level frames. Using LDPC to protect polynomial frames would again require packing into bit streams and would not preserve ring semantics across compute stages. Fountain codes, including LT and Raptor~\citep{Shokrollahi2006Raptor, Luby2002LT}, provide rate-less protection for multi-cast and high-loss environments. They are effective when receivers experience different loss patterns; however, like LDPC, they treat data as unstructured bits. They are a poor fit for our goal of keeping data algebra-compatible across compute stages.

\subsubsection{Codes over rings}
There is a long history of codes over $\mathbb{Z}_{2^k}$ for word-oriented buses. Repeated-root cyclic and negacyclic codes~\citep{DinhLopezPermouth2004ChainRings, Castagnoli1991RepeatedRoot} provide controlled redundancy and allow “derivative-style” syndrome evaluation at special points. In concrete terms, a system can add $2t$ symbols of redundancy to an $N$-symbol frame where each symbol is a $k$-bit word, then form a small number of word-level syndromes on receive to correct up to $t$ corrupted symbols. This matches fault modes like DRAM word flips or cache-line nibble errors better than bit-level designs~\citep{Schroeder2009DRAMWild, Kim2007TwoDECC}. Separately, practical encoders can be built by lifting binary BCH generators to $\mathbb{Z}_{2^k}$ and using idempotents as projectors. The concrete benefit is operational: the encoder is a ring homomorphism implemented as a single ring multiplication, so linear stages keep a protected frame inside the code by construction. When downstream logic multiplies frames, the idempotent-based encoder also preserves membership, avoiding re-encoding. Our two code families adopt precisely these concrete choices: a repeated-root negacyclic construction for $N$ that is a power of two, and a lifted BCH with an idempotent encoder for odd $N$.

\subsubsection{Interleaving.}
Production stacks already interleave to defeat bursty media: 802.11 interleaves coded bits across OFDM subcarriers~\citep{IEEE80211a1999Interleaver}; 5G NR performs bit interleaving in rate matching and then maps modulation symbols across resource elements~\citep{TS38212, TS38211}; SSD controllers interleave and stripe I/O across channels, dies, and planes (e.g., way interleaving and plane-level parallelism)~\citep{Agrawal2008SSD, Chen2011HPCA}; and RAID controllers stripe across disks~\citep{Patterson1988RAID}. These are byte/bit-level permutations that require side information or fixed layouts. For polynomial frames, we use a ring automorphism that deterministically reindexes coefficients in place. Concretely, senders apply the permutation before serialization and receivers invert it after deserialization. The effect is the same as classical interleaving—turn a cache-line, packet, or sector burst into well-spaced word errors for the decoder—without extra metadata and without leaving the algebra that later stages expect.

\subsection{Systems-Level Integrity and Recovery}

\subsubsection{Detection and Retry.}
Transmission Control Protocol (TCP) uses an end-to-end checksum and requests retransmission on loss or checksum failure~\citep{RFC793}. QUIC, a modern transport over UDP, authenticates each packet using Authenticated Encryption with Associated Data (AEAD) and performs explicit loss detection with retransmission. Storage stacks commonly attach per-block cyclic redundancy checks(CRCs); for example, HDFS verifies block checksums on read and refetches from another replica upon a mismatch~\citep{HDFS-Arch}. End-to-end file systems such as ZFS maintain object-level checksums and periodically scrub data to detect and repair latent sector errors~\citep{ZFS-FAST10}. These mechanisms are concrete and low-overhead but primarily detection-first: on any integrity mismatch, they trigger retransmission, refetch, or reconstruction. In multi-hop data paths (NICs, switches, storage, accelerators), such retries induce queuing and head-of-line blocking, amplifying tail latency even when corruption events are rare~\citep{Dean2013Tail}.

\subsubsection{Replication and storage.}
Replication and storage-level erasure coding protect durability against device or node loss after data is written. Early systems favored three-way replication (e.g., GFS/HDFS) for simplicity and fast repair~\citep{Dhulavvagol2023HDFSFederation}, while modern object stores widely use $(k+r)$ Reed–Solomon codes to cut capacity overheads. To reduce cross-rack repair traffic and recovery time, operators deploy locality-aware and bandwidth-efficient codes such as LRC, XORBAS, Hitchhiker, and Clay~\citep{Huang2012LRCSlides, Sathiamoorthy2013Xorbas, Rashmi2014Hitchhiker, Vajha2018ClayCodes}. These mechanisms are optimized for durability and efficient repair of persisted objects; they act at rest or during repair reads, not during computation. They do not prevent or correct in-flight silent corruptions on frames moving between kernels or across NICs/accelerators. As a result, a frame flipped in transit can still be consumed by the next stage unless an online check rejects it and triggers a clean replay.

\subsubsection{Fault tolerance during computation}
Algorithm-based fault tolerance (ABFT) adds checksums to matrix tiles to detect and correct arithmetic faults in GEMM-like kernels~\citep{HuangAbraham1984ABFT}; coded computation injects redundancy into task graphs to mask stragglers and node failures (e.g., polynomial codes for distributed matrix multiply)~\citep{Yu2017PolynomialCodes}. These techniques raise throughput or availability during execution, not during transport, and they operate at the kernel level, assuming control over the compute schedule. In contrast, our layer targets the boundary between kernels, protecting frames as they move across the NIC, storage, and accelerator without requiring retransmissions or kernel restructuring. Commodity hardware already corrects certain local faults: DDR memories commonly use SECDED ECC and, in DDR5, on-die ECC within the DRAM device~\citep{SemiEngineeringECC, JEDEC_DDR5}; PCIe links protect TLPs with a link CRC (LCRC) and replay unacknowledged packets via ACK/NAK at the Data Link layer~\citep{PCISIG_Basics15, Intel_PCIe_DLL}; and NVMe name-spaces can enable end-to-end data protection using Protection Information fields (Types 1–3)~\citep{NVMeCS_1_1_2024, NVMeBase_2_2_2025}. These mechanisms are essential but scoped to a single hop; they neither preserve the algebraic intent of application frames nor follow the frame across hops, so residual silent corruptions can still reach the next compute stage.

\section{Preliminary}
\subsection{Notation}
\label{sec:prelim-min}
Frames are vectors of length $N$ over $\mathbb{Z}_{2^k}$, equivalently polynomials $m(X)=\sum_{j=0}^{N-1} m_j X^j$ in the ring $R=\mathbb{Z}_{2^k}[X]/(X^N{+}1)$. If the code adds $2t$ parity symbols, the overhead is $2t$ symbols and the rate is $R_{\text{code}}=1-2t/N$. We consider $\tau$ unknown symbol errors and $\rho$ flagged erasures; bounded-distance decoding succeeds whenever $2\tau+\rho\le 2t$. To disperse bursts without changing this rule, we may apply the ring automorphism $\sigma_a(X)=X^a$ with $\gcd(a,2N)=1$. Let $M(N)$ denote one ring multiplication in $R$ (FFT/NTT: $M(N)=\Theta(N\log N)$). Decoder costs reported count the dominant $O(tN)$ word operations and omit lower-order $O(t^2)$ solves.

\subsection{Polynomial Ring Structure}
Throughout, we work in the polynomial ring
\(
R\;=\;\mathbb{Z}_{2^{k}}[X]\big/\!\big(X^{N}{+}1\big),
\)
where $N\ge 1$ and $k\ge 1$. Every element has a unique representative
\(u(X)=\sum_{j=0}^{N-1} u_j X^j\) with coefficients \(u_j\in\mathbb{Z}_{2^{k}}\),
which we identify with the length–$N$ vector of coefficients
\(
\mathbf{u}=(u_0,\ldots,u_{N-1})\in (\mathbb{Z}_{2^{k}})^{N}.
\)
Addition is coefficient-wise:
\(
u(X){+}v(X)=\sum_{j}(u_j{+}v_j)X^j,
\)
(mod $2^{k}$). Multiplication is negacyclic convolution:
\[
u(X)\cdot v(X)\ \equiv\ \sum_{j=0}^{N-1}\Big(\sum_{i=0}^{N-1} u_i\,v_{j-i\;\mathrm{mod}\;N}\,\eta(j{-}i)\Big)X^{j}\ \ (\bmod\ 2^{k}),
\]
where $\eta(t)=1$ if $0\le t\le j$ and $\eta(t)=-1$ when wrapping past degree $N{-}1$
captures the reduction $X^{N}\equiv -1$. 
so that $X$ acts as the negacyclic shift
\(
X\cdot (u_0,\ldots,u_{N-1})=(-u_{N-1},u_0,\ldots,u_{N-2}).
\)

\subsection{BCH Codes and Hensel Lifting}
\label{sec:prelim-bch}
We encode frames inside the same ambient ring $R=\mathbb{Z}_{2^k}[X]/(X^N{+}1)$. A binary BCH code of designed distance $\ delta=2t+1$ is first chosen at length $N$ over $\mathbb{F}_2$; it guarantees correction of up to $t$ unknown symbol errors, or any mix of errors/erasures satisfying $2\tau+\rho\le 2t$. To use the code at modulus $2^k$, we Hensel-lift its defining polynomials from $\mathbb{F}_2[X]$ to $\mathbb{Z}_{2^k}[X]$ so that the lifted generator still divides $X^N{+}1$ in $\mathbb{Z}_{2^k}[X]$ and defines the same BCH designed distance at the higher modulus. Operationally, this gives a structure-preserving submodule $C\subseteq R$ whose elements behave like “BCH codewords” but now carry $k$-bit symbols. 

\subsection{Negacyclic codes}
Negacyclic codes are a classical family of linear block codes built on the simple rule that a one-step rotation of a codeword flips the sign of the wrapped symbol. Introduced in the early coding-theory literature as a close cousin of cyclic and constacyclic codes, they have been used for reliable storage and communication since they admit compact algebraic descriptions and fast implementations via shift registers and FFT/NTT-style polynomial arithmetic. The following is the standard definition. Let $A$ be a commutative ring and $N \ge 1$. The negacyclic ring of length $N$ over $A$ is $R_{A,N}\ :=\ A[X]/(X^{N}+1)$. Via the identification $A^{N}\cong R_{A,N}$, we have the map
\[
(u_0,\ldots,u_{N-1})\longleftrightarrow u(X)=\sum_{j=0}^{N-1}u_j X^j\ \ (\bmod\ X^{N}+1),
\]
the relation $X^{N}\equiv -1$ induces the negacyclic shift $\tau(u_0,\ldots,u_{N-1})\ =\ (-u_{N-1},\,u_0,\,\ldots,\,u_{N-2})$. An $A$–linear code $\mathcal C\subseteq A^{N}$ is called negacyclic if it is closed under $\tau$; equivalently, under the above identification, $\mathcal C$ corresponds to an $A$–submodule of $R_{A,N}$.
When $A$ is a field, negacyclic codes are precisely the principal ideals
$\langle g(X)\rangle\subset R_{A,N}$ with $g(X)\mid X^{N}{+}1$ in $A[X]$. In particular, when $N$ be odd, over the residue field $\mathbb F_{2}$, $\bar R\ :=\ \mathbb F_{2}[X]/(X^{N}+1)$ is semisimple, i.e., $X^{N}+1$ is square-free in $\mathbb F_{2}[X]$ and $\bar R$ is isomorphic to $\prod_{i=1}^{s}\mathbb F_{2^{d_i}}$ by the Chinese remainder theorem. However, when $N$ is a power of two, the binary residue is no longer semisimple. In that case, over $\mathbb{F}_2$ one has $X^{N}+1=(X+1)^{N}$, therefore
\[
\bar R\;:=\;\mathbb{F}_2[X]/(X^{N}+1)\;\cong\;\mathbb{F}_2[Y]/(Y^{N})\quad\text{with }Y:=X+1.
\]
Which is a local ring where $Y$ is nilpotent. Therefore $R$ is local with maximal ideal $\langle 2,\,Y\rangle$ and residue field
$R/\langle 2,\,Y\rangle\cong\mathbb{F}_2$; in particular, an element $u\in R$ is a unit
iff $u(-1)\bmod 2\neq 0$. The principal ideals $\langle Y^{r}\rangle=\big\langle (X+1)^{r}\big\rangle\subset R$ where $0\le r\le N$, form a descending chain
$R\supset\langle Y\rangle\supset\cdots\supset\langle Y^{N}\rangle=\{0\}$.
A basic divisibility test that we will use is: $Y^{r}\mid f(X)\ \Longleftrightarrow\ f^{[i]}(-1)=0$ for all $\,0\le i<r$,
where $f^{[i]}$ denotes the $i$-th Hasse derivative.
In particular, the repeated-root negacyclic codes
\(
\mathcal{C}_{t}:=\langle (X+1)^{2t}\rangle\subset R
\)
are ideals closed under addition and multiplication by ring elements, and admit
systematic encoding $c(X)=m(X)(X+1)^{2t}\bmod (X^{N}+1)$ with $\deg m<N-2t$.

\section{System Workflow}
\label{sec:system}
In many modern data pipelines, data are represented as polynomials over cyclotomic quotient rings and must move across processors, accelerators, and storage tiers. Embedding vectors into polynomials enables FFT/NTT-based transforms that reduce computation from $O(n^{2})$ to $O(n\log n)$ for length-$n$ inputs. This section surveys reliability strategies for such polynomial-frame workflows—spanning streaming analytics, coded computation, signal processing, and scientific simulation—and contrasts them with our approach. We first review two widely used baselines, hash-and-check and plain bytewise BCH, and explain why they are ill-suited for ring-aware pipelines that require structure preservation. We then present a ring-compatible error-correcting layer with two concrete instantiations and describe how it composes with polynomial-encoded circuits and operators. Finally, we formalize the fault model, derive decoding guarantees, and briefly discuss security considerations relevant to real deployments.

\subsection{Baseline Model}
\begin{figure}[t]
  \centering
      \includegraphics[width=0.7\linewidth,trim=0cm 14cm 0cm 1cm,clip]{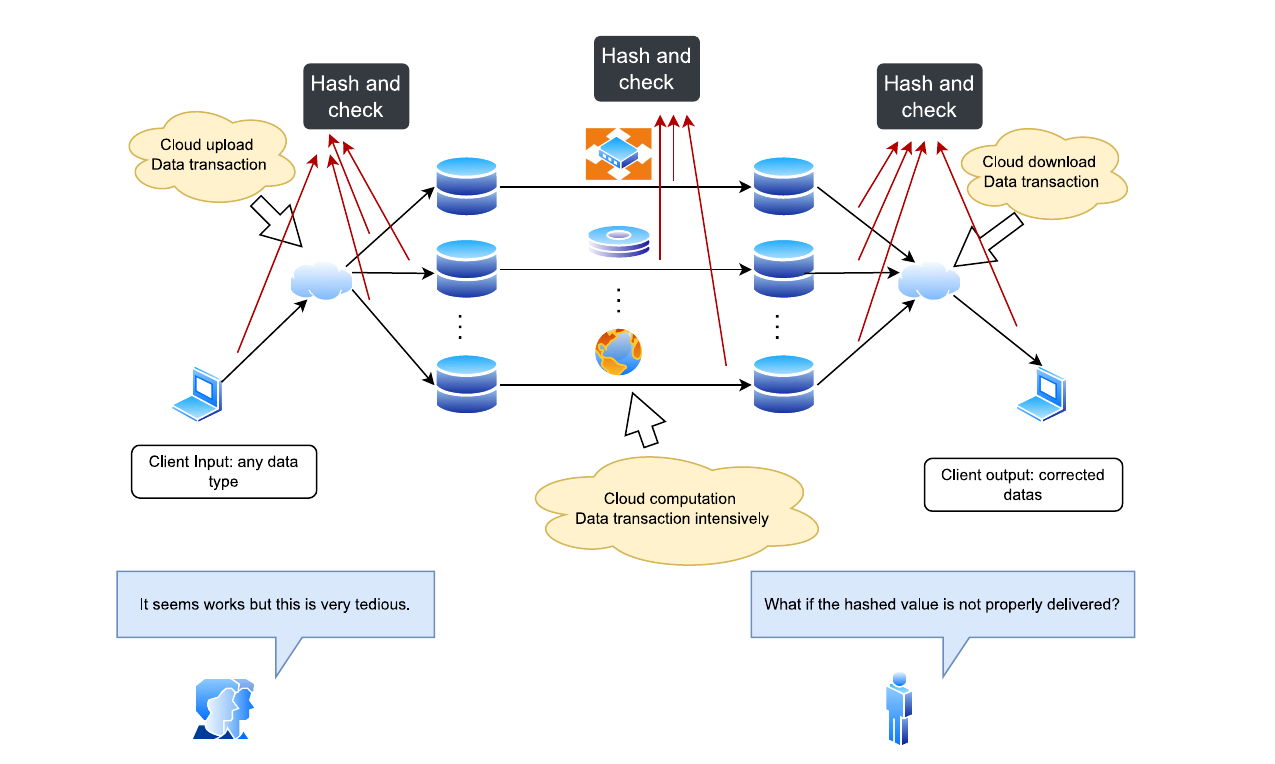}
  \caption{Hash-and-check workflow: interactive detection only, RTT on the critical path.}
  \label{fig:trivial}
\end{figure}

\begin{figure}[t]
  \centering
   \includegraphics[width=0.7\linewidth,trim=0cm 13cm 0cm 1cm,clip]{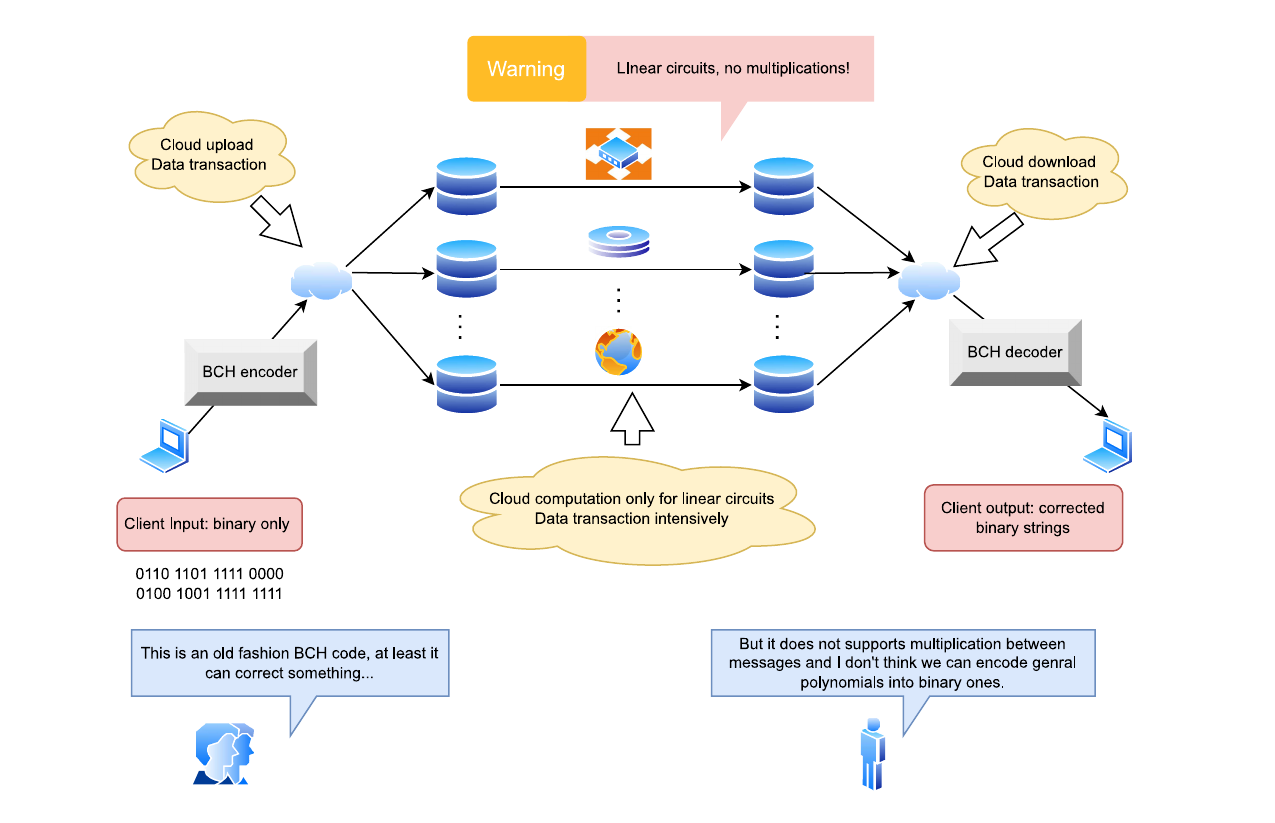}
  \caption{Plain BCH on bitstreams: good for transport, but not ring-compatible with HE.}
  \label{fig:bch}
\end{figure}

A common baseline for reliability regarding polynomially encoded data is hash-and-check: namely, after each compute stage, the admin for that process hashes the output vector and runs an interactive verification protocol with the admin of the incoming stage. If they cannot agree on a hashed value, they have to resend the data and redo the hash-and-check process. Another baseline is a plain binary BCH code over binary data applied to bit-strings before transporting. This construction toward reliability is somewhat non-interactive since BCH has an algebraic property that can self-detect and correct a couple of bit flips after a binary message is corrupted by some noise. But due to its incompatibility when encountering multiplication circuits, the admin needs to interactively check with the coming computation stage unless the circuit is linear. Both of the above have distinct drawbacks for HE pipelines:

\begin{itemize}
\item \textbf{Hash-and-check.}
Figure~\ref{fig:trivial} shows the workflow of the hash-and-check attempt. Each hop requires a round trip of at least hashed value, and any failures
stall on the network round-trip time (RTT). Further, verification only detects corruption without correcting and forces
retransmission or re-execution, which is costly in latency-sensitive settings, e.g., privacy-preserving ML inference. Hash functions are also fragile to in-flight bit flips: if the hash value is corrupted by SDC, the check fails spuriously. In addition, this approach requires a very intensive interactive verification process; thus, it is slow, and the system is fully deactivated even with one corrupted database.

\item \textbf{Plain BCH.}
Binary BCH codes are defined over $\mathbb{F}_2$ and protect bit-strings well with an efficient BCH encoder,
but most HE computation is performed on ring elements in $R=\mathbb{Z}_{2^k}[X]/(X^N\!\pm\!1)$.
Mapping between these domains breaks ring semantics, since there is no isomorphism between
different characteristics in general, which prevents the preservation of addition/multiplication
through the reliability layer. In practice, a binary outer code constrains the
workflow to linear operations on encoded payloads and introduces packing
and unpacking overheads; see Figure~\ref{fig:bch} for a concrete workflow. Further, BCH encoding causes the message length to be longer and, therefore, there is a higher probability for more bit-flips than planned, which breaks the reliability of the BCH layer. This parameter tuning requires more careful planning and may destroy the reliability of the BCH code.
\end{itemize}

\subsection{Our Approach}
\begin{figure}[t]
  \centering
  \includegraphics[width=0.7\linewidth,trim=0cm 14cm 0cm 1cm,clip]{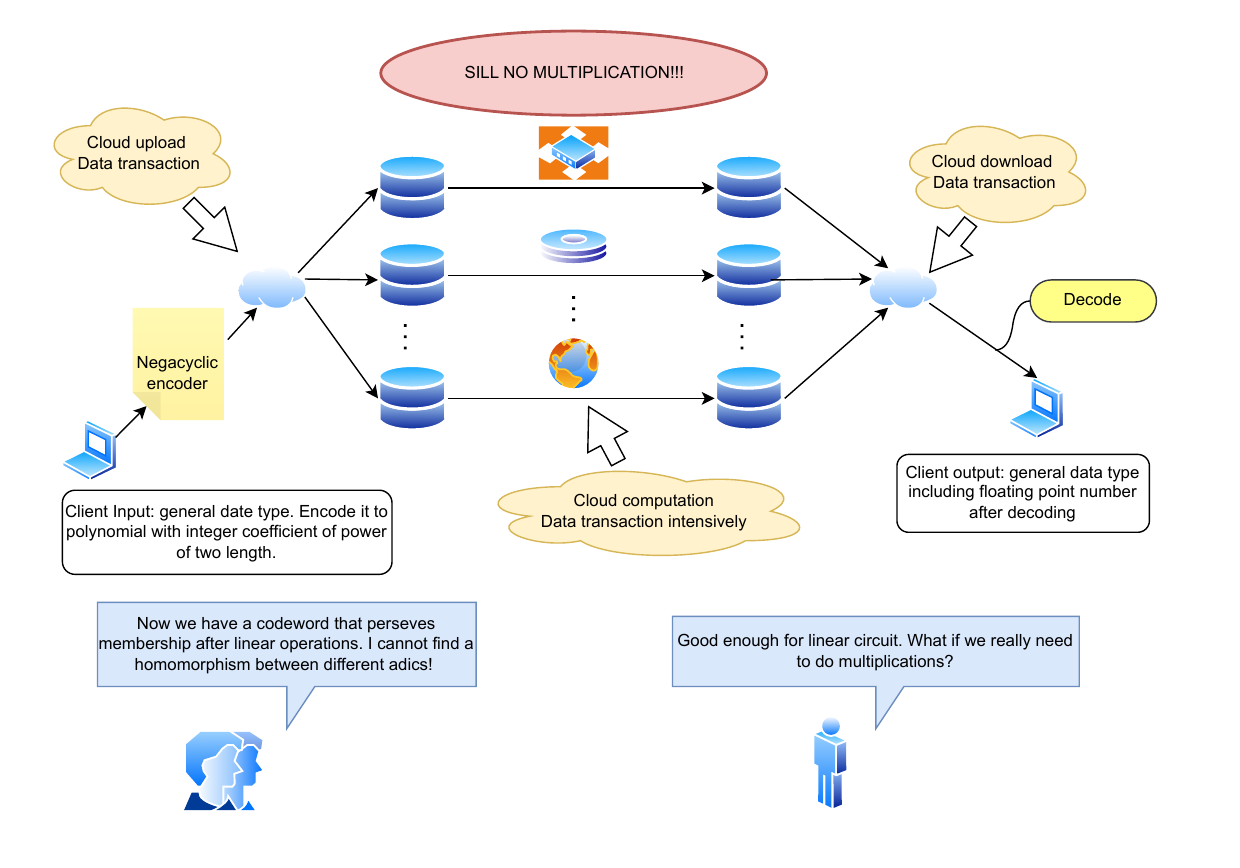}
  \caption{Repeated-root: general input domain, linear HE preserved.}
  \label{fig:liftbch}
\end{figure}

\begin{figure}[t]
  \centering
  \includegraphics[width=0.7\linewidth,trim=0cm 14cm 0cm 1cm,clip]{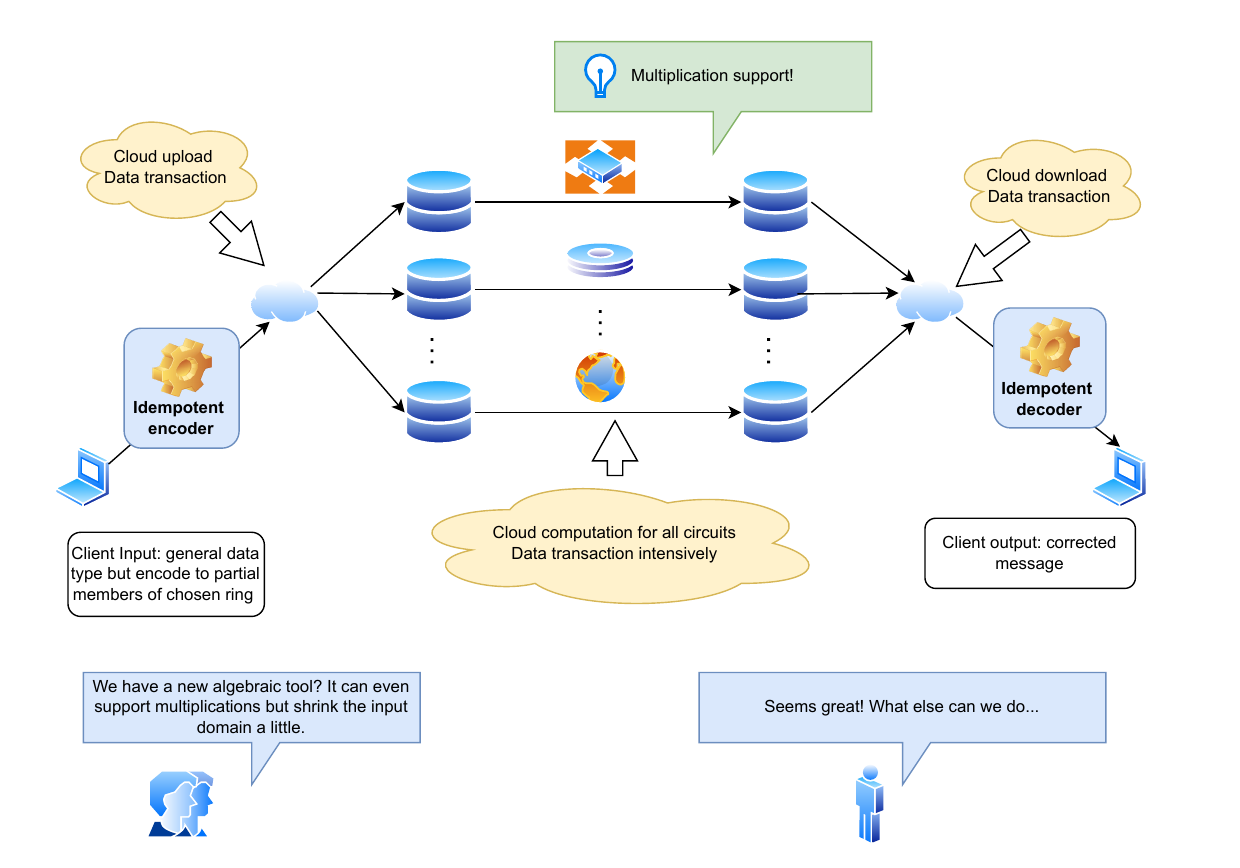}
  \caption{Idempotent encoder: shrinks plaintext domain to $Re$ but preserves add \emph{and} mult.}
  \label{fig:idemp}
\end{figure}

We design an error-correcting layer that lives in the same ring as the
polynomial frames pipeline payload and preserves the needed algebra. It is non-interactive, corrects errors in flight, and composes with HE evaluation naturally. We can also trade off the data rate for less frequent error correcting. Ideally, a one-time error correcting at the end can sufficiently correct all errors, and this process can even be made fully offline, but this would also require careful parameter tuning. We provide two constructions as their workflows are shown in Figure~\ref{fig:liftbch} and Figure~\ref{fig:idemp}.

\begin{enumerate}
\item \textbf{Repeated-root Negacyclic Code.}
For $R=\mathbb{Z}_{2^k}[X]/(X^{2^m}+1)$, we use the ideal
$\mathcal{C}_t=\langle(X+1)^{2t}\rangle$ to generate the codeword. Encoding is systematic:
$c(X)=m(X)(X+1)^{2t}\bmod(X^{2^m}+1)$. The codeword is closed under addition and
multiplication by ring elements, in particular, the product of two encoded messages is still encoded, but we cannot ensure $Enc(a)Enc(b) = Enc(ab)$. However, linear HE circuits do preserve
code membership and circuit operations, which induce potential applications in privacy-preserving linear programming. Decoding uses Hasse-derivative syndromes at $X=-1$ and corrects up to $t$ symbol errors over $\mathbb{Z}_{2^k}$ with $O(tN)$ syndrome cost and $O(t^2)$ locator/magnitude recovery cost for small $t$ based on the selection of codeword.

\item \textbf{Lifted BCH with Idempotent Encoder.}
For $R=\mathbb{Z}_{2^k}[X]/(X^N+1)$ with $N$ odd, we lift a binary BCH generator
$g_2$ to $\tilde g$ by Hensel lifting, and take $\mathcal{C}=\langle\tilde g\rangle=Re$,
where $e$ is the CRT idempotent. Encoding $m\mapsto me$, equivalently, $m\tilde g$ is a ring homomorphism that preserves addition and multiplication in all cases, i.e., $Enc(a)Enc(b) = Enc(ab)$, so general HE circuits
preserve code membership. The residue code has designed distance $\delta=2t{+}1$;
decoding corrects $t$ symbol errors, depending on the selection of the BCH code to be lifted, via binary BCH on residues plus $2$-adic lifting
of error magnitudes.
\end{enumerate}

\subsection{System and Fault Model}
\label{subsec:model}
In this subsection, we formalize the system fault model and prove that our construction can efficiently defeat SDC, including symbol flips and bursts. As aforementioned, we protect ring elements in $R=\mathbb{Z}_{2^k}[X]/(X^N+1)$ in polynomial/vector form, where  $N\in\{2^m\}\ \text{or}\ N\ \text{ is odd}$ 
and, considering real-world deployments, $k\in\{8,16,32\}$, i.e., byte/halfword/word symbols (we can extend the construction to larger $k$ trivially). Concretely, a codeword
$c(X)=\sum_{j=0}^{N-1} c_j X^j\in R$ in our system is produced
by a systematic encoder:
\begin{enumerate}
\item Repeated-root when $N=2^m$: $c(X)=m(X)(X+1)^{2t}\bmod(X^N+1)$ with
$\deg m<N-2t$, code $\mathcal{C}_t=\langle(X+1)^{2t}\rangle$, redundancy $2t$
coefficients.
\item Lifted BCH when $N$ Odd: $c(X)=m(X)\tilde g(X)\bmod(X^N+1)$ with
$\tilde g$ the Hensel lift of a binary BCH generator, code $\mathcal{C}=\langle\tilde g\rangle=Re$.
\end{enumerate}
We transmit/store the coefficient vector $(c_0,\dots,c_{N-1})$ in little-endian
order where the least significant byte (LSB) is placed at the lowest memory address and each coefficient is a symbol in $\mathbb{Z}_{2^k}$ (i.e., $k$-bit word). We define our error model as the following: 
a channel/storage fault corrupts a codeword by adding an error polynomial
$E(X)=\sum_{j=0}^{N-1} e_j X^j\in R$, yielding $r(X)=c(X)+E(X)\ \in R$, and $e_j\in\mathbb{Z}_{2^k}$. We say position $j$ is a symbol error if $e_j\neq 0$. The Hamming weight
$\mathrm{wt}(E):=|\{j: e_j\neq 0\}|$ is the number of erroneous symbols.
We consider two processes:

\begin{enumerate}
\item \emph{i.i.d.\ symbol flips:} Each symbol position $j\in\{0,\dots,N-1\}$ is corrupted independently with probability $p$. That is, $\Pr[e_j\neq 0]=p$ and $\Pr[e_j=0]=1-p$. When a symbol is corrupted, its nonzero value in $\mathbb{Z}_{2^k}\setminus\{0\}$ may follow any distribution; our decoders are designed to handle arbitrary nonzero magnitudes. Under this model, the total number of corrupted symbols satisfies $\mathrm{wt}(E)\sim \text{Binomial}(N,p)$.

\item \emph{Burst errors:} A contiguous run of $B$ symbol positions is corrupted. That is, for some starting index $j_0$, the symbols $e_{j_0}, e_{j_0+1}, \dots, e_{j_0+B-1}$ (interpreted modulo $N$) are all nonzero, while all other positions are error-free. This model localized corruption within a packet, cacheline, or storage sector, where faults affect consecutive symbols in memory or transmission order.

\end{enumerate}

Optionally, lower layers—such as a per-packet cyclic redundancy check (CRC)—may flag a subset of indices $\mathcal{E} \subseteq \{0,\dots, N-1\}$ as erasures. These locations are known to be suspect, while any remaining corruptions are treated as random errors. Our decoder supports the standard error/erasure tradeoff, correcting any pattern satisfying $2\tau + \rho \le 2t$. To combat burst errors, we apply a ring automorphism $\sigma_a: R \to R$, defined by $X \mapsto X^a$, where $a$ is an odd integer coprime to $2N$ (see Section~\ref{sec:automorphism-interleave}). This automorphism permutes coefficients via the index map $j \mapsto a j \mod N$, which is a bijection. We apply $\sigma_a$ before transmission and its inverse $\sigma_{a^{-1}}$ after reception. A burst of $B$ consecutive indices, starting at $j_0$, is mapped to the set $\{aj_0, a(j_0+1), \dots, a(j_0 + B - 1)\} \bmod N$. This results in $B$ symbols distributed as an arithmetic progression with stride $a$. When $a$ is chosen uniformly at random from the $(\varphi(2N)/2)$ odd units, the resulting indices are nearly uniform over $\{0,\dots,N-1\}$. Collisions only arise due to wrap-around effects when the burst length $B$ becomes comparable to $N$.

Our decoding strategy in this systemic manner: 
\begin{enumerate}
\item Repeated-root $N=2^m$: the Hasse-derivative decoder corrects any
$E$ with $\mathrm{wt}(E)\le t$; with erasures $\rho:=|\mathcal{E}|$ known and
$\tau$ unknown errors, correction is guaranteed when $2\tau+\rho\le 2t$.
\item Lifted BCH $N$ Odd: the binary residue code has designed distance
$\delta=2t+1$; we decode any $E$ with $\mathrm{wt}(E)\le t$. With erasures,
the usual bound $2\tau+\rho<\delta$ applies. Error magnitudes are then
recovered modulo $2^k$ via $2$-adic lifting, which induces a unique solution since the relevant.
Vandermonde determinants are odd units.
\end{enumerate}

\subsubsection{Reliability Check}
Under the i.i.d.\ error model, let $P_{\mathrm{fail}}$ denote the block failure probability—i.e., the probability that the decoder either fails to produce a result or outputs an incorrect codeword. For i.i.d.\ symbol errors with rate $p$, the error-only case satisfies
\[
P_{\mathrm{fail}}^{\text{iid}} = \Pr[\mathrm{wt}(E) > t] = \sum_{j = t+1}^{N} \binom{N}{j} p^j (1 - p)^{N - j}.
\]
This expression gives the tail probability of a Binomial$(N, p)$ distribution exceeding the decoding radius $t$. If $\rho$ erasures are also flagged—chosen uniformly at random and independently of symbol errors—then decoding succeeds when $2\tau + \rho \le 2t$. The corresponding failure probability is the total weight of the joint binomial tails over all pairs $(\tau, \rho)$ that violate this inequality. This accounts for the combined impact of errors and erasures on the decoder's success. 

\begin{figure}[t]
  \centering
      \includegraphics[width=0.9\linewidth]{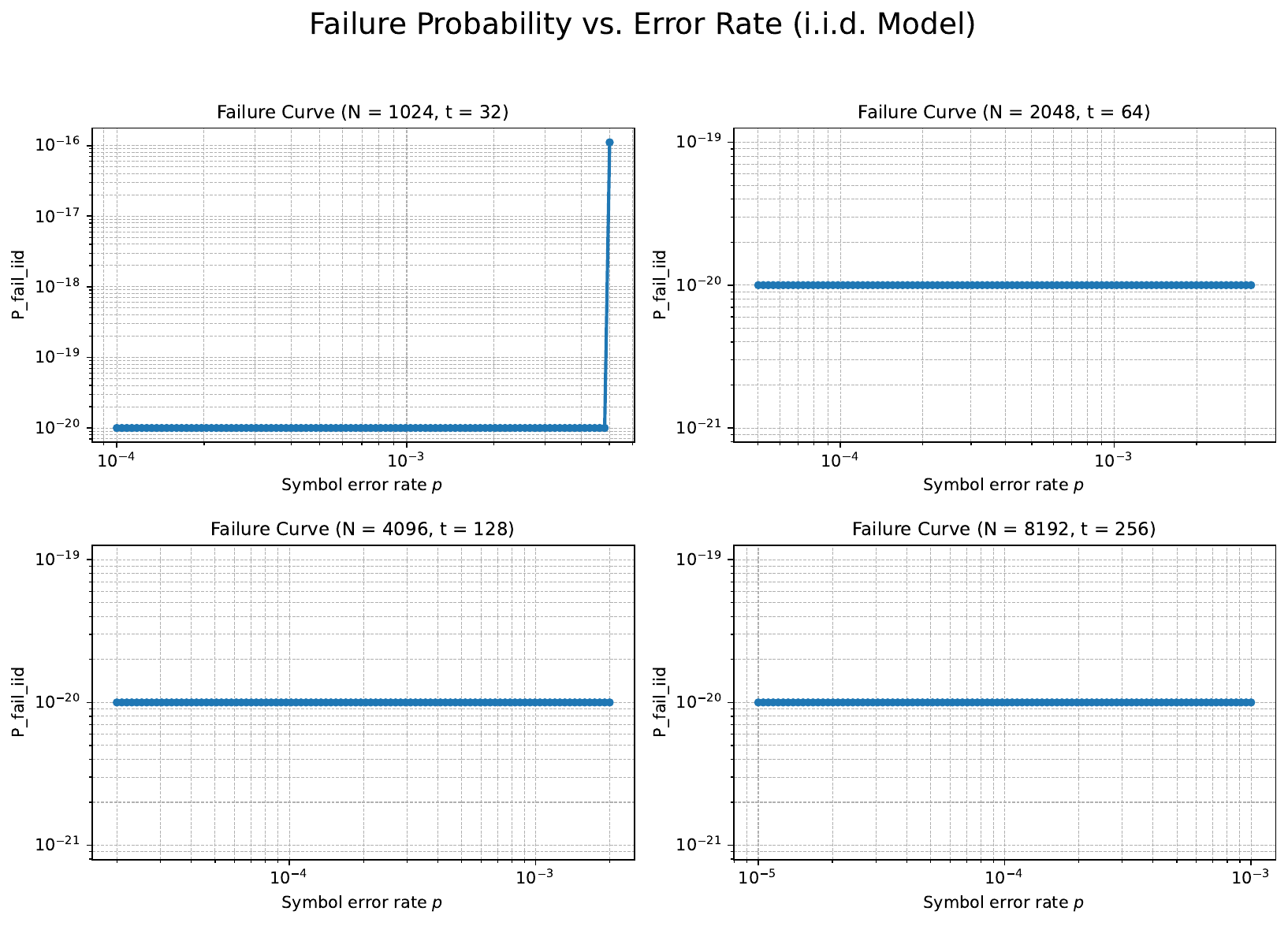}
    \caption{Block failure probability under i.i.d.\ symbol flips, shown for $N \in \{1024, 2048, 4096, 8192\}$ with $t = N/32$. Each curve captures the sharp transition as $p$ increases. The $N{=}1024$ case shows a discrete jump due to binomial granularity at small $t$.
    }
  \label{fig:failiid}
\end{figure}

We plot the block failure probability $P_{\mathrm{fail}}^{\mathrm{iid}} = \Pr[\mathrm{wt}(E) > t]$ 
under the i.i.d.\ symbol error model as in Figure~\ref{fig:failiid}. 
Four subplots visualize this failure probability for increasing block lengths 
$N \in \{1024, 2048, 4096, 8192\}$, each with decoder radius $t = N/32$. For each case, we carefully choose a range of $p$ values based on expected SDC rates, and the failure rates are negligible. The apparent outlier in the $N = 1024$ plot is due to a numeric discontinuity when $\Pr[\mathrm{wt}(E) > t]$ transitions sharply across just a few integer values of error count $j > t$, a behavior more pronounced when $N$ and $t$ are small. As $N$ increases, the binomial distribution smooths out and the curves appear more continuous.

For the model under bursts with interleaving, let a single burst of length $B$ occur. After interleaving by $\sigma_a$, the
$B$ affected positions are separated by stride $a$ modulo $N$; if $a$ is chosen
uniformly among the odd units, the indicators $\mathbf{1}\{e_j\neq 0\}$ seen by
the decoder are close to i.i.d.\ Bernoulli with mean $\approx B/N$. A simple and useful bound is
\[
\mathbb{E}[\mathrm{wt}(E)]\;=\;B,\text{ and }
\mathrm{Var}[\mathrm{wt}(E)]\;\le\;B\!\left(1-\frac{B}{N}\right),
\]
and the block error probability can be bounded by a Chernoff tail around mean $B$ with threshold $t$; in practice, choosing $a$ per-frame from a public seed and
$B\ll N$ yields dispersion sufficient to meet $t$ with high probability. Multiple
disjoint bursts superpose linearly in this analysis. If a lower layer flags the
burst region as erasures (e.g., packet loss), the decoder succeeds under $2\tau+\rho\le 2t$ (resp.\ $<\delta$). 

Particularly, in one of the most common polynomial frames pipelines: FHE, such as BFV/BGV with plaintext modulus $t=2^k$, the plaintext space is exactly $\mathbb{Z}_{2^k}[X]/(X^N\!\pm\!1)$, so the reliability layer can wrap plaintexts
or linear-intermediate results in place. For the PPML application, i.e., CKKS, if the application uses
quantized fixed-point with scale $\Delta$ and keeps the numerical error below
$\Delta/2$, rounding to $\mathbb{Z}_{2^k}$ produces exact symbols that the layer
can protect; otherwise, the layer protects transport but cannot remove inherent
approximation. For ciphertexts in RNS form where modulus $Q=\prod_i q_i$, our layer
is applied outside the HE ciphertext algebra (i.e., after serialization
into $\mathbb{Z}_{2^k}$ words), leaving RLWE noise and key material unchanged. However, to generate a new correcting layer and link the error correcting property to each layer is complicated and has not yet been solved. Based on our knowledge, there exists research on non-level CKKS schemes, and those instantiations offer a suitable environment for our current error-correcting layer.

\subsection{Security Considerations}
\label{subsec:security}

We assume that confidentiality and integrity are provided at the session layer by a cryptographic AEAD scheme. Our error-correcting code (ECC) layer is public, deterministic, and not intended to protect secrets or reduce homomorphic noise. It targets benign in-flight corruption of polynomial frames, such as soft memory faults or silent data corruption (SDC), during transport or storage.

To preserve standard notions such as IND-CPA and INT-CTXT, ECC encoding must operate under message authenticity. On send, the sender first computes $c \leftarrow \mathrm{Enc}(m)$, then wraps it in an authenticated encryption envelope: $\tilde{c} \leftarrow \mathrm{AEAD.Enc}(K, c, {\sf hdr})$. On receive, decoding must occur in constant time: extract the payload from $\tilde{c}$, run ECC decoding to recover $c$, verify the AEAD tag, and only then release the plaintext $m \leftarrow \mathrm{AEAD.Dec}(K, c, {\sf hdr})$. No observable behavior (e.g., success/failure) should be exposed prior to tag verification. Decoder implementations must run in constant time with respect to corrupted input. They must avoid data-dependent branching, memory access patterns, or variable-time loops. When $\tau > t$, bounded-distance decoders may miscorrect; to detect such residual errors, a short CRC over the frame (or equivalently, AEAD failure) is required. When erasures are available, apply the standard condition $2\tau + \rho \le 2t$ to reduce miscorrection risk.

The automorphism interleaver $\sigma_a: X \mapsto X^a$ is algebra-preserving and stateless. It does not require a key and maintains both the ring and code structure. The parameter $a$ should be chosen to have a long cycle length and fixed per session or flow. This prevents adversarial alignment of bursts that might defeat interleaving. ECC must not be applied to secret material, such as secret keys or switching/relinearization keys; doing so would introduce algebraic structure over secret values, which could be exploited by attackers.
\subsection{End-to-End Instantiation}
\label{subsec:workflow}
\begin{figure}[t]
\centering
\includegraphics[width=0.9\linewidth,trim=0cm 18cm 0cm 1cm,clip]{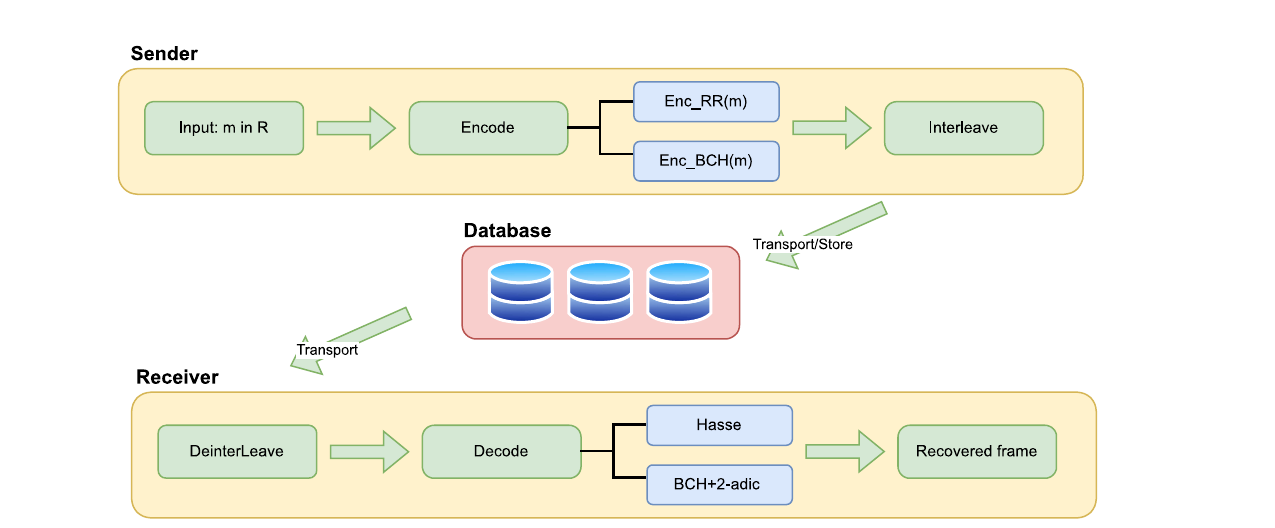}
\caption{End-to-end workflow}
\label{fig:workflow}
\end{figure}

Let $R=\mathbb{Z}_{2^k}[X]/(X^N+1)$ with $N\in\{2^m\}$ or $N$ odd. A \emph{frame} is a polynomial $m(X)=\sum_{j=0}^{N-1} m_j X^j\in R$ with coefficients
$m_j\in\mathbb{Z}_{2^k}$. We have the encoding function aforementioned
write $\tilde g$ the Hensel lift of a binary BCH generator. The automorphism interleaver
is $\sigma_a:X\mapsto X^a$ with $a$ odd and $\gcd(a,2N)=1$; its inverse uses $a^{-1}\!\!\pmod{2N}$. We optionally attach a CRC word to enable erasure flags. The sender follows Algorithm~\ref{alg:sender} and the receiver follows Algorithm~\ref{alg:receiver}. Figure~\ref{fig:workflow} visualizes this instantiation. 

\begin{algorithm}[H]
\small
\caption{Sender: \textsc{EncodeAndSend}$(m, \mathtt{mode}, t, a)$}
\label{alg:sender}
\begin{algorithmic}[1]
\Require $m\in R$; $\mathtt{mode}\in\{\mathrm{RR},\mathrm{BCH}\}$; $t\ge 1$; odd $a$ with $\gcd(a,2N)=1$
\Ensure One-way transmission of interleaved codeword $c'$
\If{$\mathtt{mode}=\mathrm{RR}$} \Comment{Repeated-root ($N=2^m$)}
  \State $c \gets m(X)\cdot (X+1)^{2t}\bmod(X^N{+}1)$
\ElsIf{$\mathtt{mode}=\mathrm{BCH}$} \Comment{Lifted BCH ($N$ odd)}
  \State $c \gets m(X)\cdot \tilde g(X)\bmod(X^N{+}1)$ \hfill \small{$O(N\log N)$ with NTT}
\EndIf
\State $c' \gets \sigma_a(c)$ \hfill \small{index map $j\mapsto a j\!\!\pmod{N}$; in-place $O(N)$}
\State \textbf{send} $c'$ (and optional CRC)
\end{algorithmic}
\end{algorithm}

\begin{algorithm}[H]
\small
\caption{Receiver: \textsc{RecvAndDecode}$(c', \mathtt{mode}, t, a^{-1})$}
\label{alg:receiver}
\begin{algorithmic}[1]
\Require Interleaved word $c'$; $\mathtt{mode}\in\{\mathrm{RR},\mathrm{BCH}\}$; $t\ge 1$; $a^{-1}\equiv a^{-1}\ (\bmod\,2N)$
\Ensure Recovered frame $\hat m\in R$ or \textsc{Fail}
\State $\hat c \gets \sigma_{a^{-1}}(c')$ \hfill \small{undo interleaving; $O(N)$}
\If{$\mathtt{mode}=\mathrm{RR}$} \Comment{Hasse-derivative decoder}
  \State Compute syndromes $S_i \gets \hat c^{[i]}(-1)$ for $i=0,\dots,2t-1$  \hfill \small{$O(tN)$}
  \If{$S_i=0$ for all $i$} \Return $\hat m \gets \hat c/(X+1)^{2t}$ in $R$ \hfill \small{strip the $Y^{2t}$ factor}
  \Else
    \State Run Berlekamp--Massey/EEA on $(S_i)$ to get locator $\Lambda(z)$ of degree $\nu\le t$ \hfill \small{$O(t^2)$}
    \State Find error positions $\{j_\ell\}$ from roots of $\Lambda$ (Chien-like search) \hfill \small{$O(tN)$}
    \State Solve for magnitudes $\{e_\ell\}$ via a $\nu\times \nu$ Vandermonde system over $\mathbb{Z}_{2^k}$ \hfill \small{$O(t^2)$}
    \State $\hat c \gets \hat c - \sum_{\ell=1}^{\nu} e_\ell X^{j_\ell}$; \quad \Return $\hat m \gets \hat c/(X+1)^{2t}$ in $R$
  \EndIf
\ElsIf{$\mathtt{mode}=\mathrm{BCH}$} \Comment{Residue BCH + $2$-adic lifting}
  \State $\bar c \gets \hat c \bmod 2$ in $\mathbb{F}_2[X]/(X^N{+}1)$
  \State Decode $\bar c$ with binary BCH to correct up to $t$ positions $\{j_\ell\}$ \hfill \small{$O(t^2)$}
  \State Lift error magnitudes $e_\ell\in\mathbb{Z}_{2^k}$ by Hensel: solve triangular system modulo $2,4,\dots,2^k$
  \State $\hat c \gets \hat c - \sum_{\ell=1}^{\nu} e_\ell X^{j_\ell}$; \quad \Return $\hat m \gets \hat c/\tilde g$ in $R$
\EndIf
\State \textbf{return} \textsc{Fail} \Comment{if any step aborts (optional CRC can flag erasures)}
\end{algorithmic}
\end{algorithm}

\section{Error-Correcting Layers Construction}
\label{sec:ecc-layer}

In this section, we describe an in-ring error-correcting layer for polynomial-frame pipelines, e.g., streaming analytics, coded computation, signal processing, and scientific simulation, over rings of the form
$R=\mathbb{Z}_{2^{k}}[X]/(M(X))$, where $\mathbb{Z}_{2^{k}}$ is a ring modulus (distinct from the finite field $\mathbb{F}_{2^{k}}$~\citep{MullenPanario2013}) and $M(X)=X^{N}+1$ with $N$ the frame length. The layer preserves algebraic structure so downstream linear/multiplicative operators remain valid on coded data. We present two complementary instantiations that cover common lengths—one for powers of two and one for odd $N$—providing ring-compatible encoding/decoding with bounded-distance error correction.

\begin{enumerate}
  \item multiplicative encoder in a semisimple negacyclic ring with odd length, yielding a ring-homomorphic map whose image is an ideal code closed under $+$ and $\times$, therefore we can use this error-correcting layer for general polynomial encoded based circuits, but this encoder is not always injective, thus we need to shrink the input domain or track some property for accurate decoding.
  \item a repeated-root negacyclic ideal code for power-of-two lengths, closed under $+$ and $\times$ with efficient syndrome decoding. This means: $Enc_B(a)Enc_B(b)$ preserves error-correcting property but $Enc_B(a)Enc_B(b)\neq Enc_B(ab)$, so this encoder works with applications that only require linear operations. Such application includes general linear programming and private-preserving linear programming.  
\end{enumerate}

\subsection{When $N$ is Odd}
\label{subsec:semisimple}

Let $N\ge 1$ be odd and set $M(X)=X^{N}+1$. Working over modulo $2$, we have $M(X)\equiv X^{N}-1$, which is square-free because $N$ is odd and
$\tfrac{d}{dX}(X^{N}-1)=N X^{N-1}\not\equiv 0\pmod{2}$.
Hence $X^{N}-1=\prod_{i=1}^{s} f_i(X)$ over $\mathbb{F}_2[X]$ with distinct monic irreducibles $f_i$.
Each $f_i$ lifts uniquely to a monic, pairwise-coprime $\tilde f_i\in \mathbb{Z}_{2^{k}}[X]$ and
\begin{equation}\label{eq:lifted-factor}
M(X)\;=\;\prod_{i=1}^{s} \tilde f_i(X)\quad \text{in } \mathbb{Z}_{2^{k}}[X].\notag
\end{equation}
By CRT, we obtain a semisimple decomposition
\(
R \cong \prod_{i=1}^{s} R_i,\; R_i=\mathbb{Z}_{2^{k}}[X]/(\tilde f_i).
\) With this notation, we define idempotents in this ring as the following definition, and the properties for this algebraic object are specified by Lemma~\ref{lem:idempotent-encoder}. To find all such idempotents, we have a concrete helper via EEA as shown in Algorithm~\ref{alg:invmod-hensel}.

\begin{definition}[Primitive idempotents and ideal codes]
For each $i$, write $M_i(X)=M(X)/\tilde f_i(X)$ and let $u_i(X)$ be the inverse of $M_i(X)$ modulo $\tilde f_i(X)$.
Define the primitive idempotent $e_i \in R$ by
\(
e_i \equiv M_i(X) \cdot u_i(X)\ \ (\bmod\ M(X)).
\)
For an index set $\mathcal{I}\subseteq\{1,\ldots,s\}$, let
\(
e \;=\; \sum_{i\in \mathcal{I}} e_i
\)
and the associated ideal code
\(
\mathcal{C} \;=\; R e \;=\; \{ a e : a\in R\}.
\)
\end{definition}

\begin{lemma}[Idempotent encoder]\label{lem:idempotent-encoder}
Let $N$ be odd, $M(X)=X^N+1\in\mathbb{Z}_{2^k}[X]$, and suppose
$M=\prod_{i=1}^s \tilde f_i$ with monic, pairwise coprime $\tilde f_i\in\mathbb{Z}_{2^k}[X]$
(the Hensel lifts of the distinct irreducible factors of $X^N-1$ over $\mathbb{F}_2$).
For each $i$, set $M_i:=M/\tilde f_i$ and pick $u_i\in\mathbb{Z}_{2^k}[X]$ satisfying
$M_i u_i \equiv 1 \pmod{\tilde f_i}$. Define $e_i \in R:=\mathbb{Z}_{2^k}[X]/(M)$ by
$e_i \equiv M_i u_i \pmod{M}$, and for any index set $\mathcal I\subseteq\{1,\dots,s\}$ let
$e:=\sum_{i\in\mathcal I} e_i$ and $\mathcal{C}:=Re$.
Then:
\begin{enumerate}
\item $e$ is idempotent: $e^2=e$.
\item The map $Enc:R\to R$, $Enc(m)=m e$, is a ring homomorphism with image $\mathcal C=Re$.
\item Writing $\bar{\mathcal I}:=\{1,\dots,s\}\setminus\mathcal I$ and
$h_{\bar{\mathcal I}}(X):=\prod_{j\in\bar{\mathcal I}}\tilde f_j(X)$, one has the principal ideal identity
\[
\mathcal C \;=\; Re \;=\; \langle\, h_{\bar{\mathcal I}}(X)\,\rangle \;\subseteq\; R.
\]
\end{enumerate}
\end{lemma}
\begin{proof}
See Appendix~\S\ref{proof lem enc}.
\end{proof}

\subsubsection{Choosing distance via a BCH generator.}
Given the fact that we can efficiently compute all primitive idempotents, how can we use them as an algebraic encoder? We consider lifting a BCH code in the following manner: Fix a designed distance $\delta=2t+1$ for a binary BCH code of length $N$. Let $\mathcal{S}\subset\{0,\ldots,N-1\}$ be the usual consecutive root-exponent set over $\mathbb{F}_2$; let
\(
g_2(X)=\prod_{i:\ f_i \text{ has roots in }\mathcal{S}} f_i(X)\in\mathbb{F}_2[X]
\)
Be the binary BCH generator. Its lift
\(
\tilde g(X)=\prod_{i:\ f_i\mid g_2} \tilde f_i(X)\in \mathbb{Z}_{2^k}[X]
\)
satisfies $\tilde g\mid M$ and generates a free cyclic code $\langle \tilde g\rangle\subseteq R$ whose residue modulo $2$ is the chosen BCH code.
Taking $\mathcal{I}=\{i:\ f_i\mid g_2\}$ and $e=\sum_{i\in\mathcal{I}} e_i$ (Lemma~\ref{lem:idempotent-encoder}) yields a \emph{multiplicative encoder} $Enc(m)=m e$ into $\mathcal{C}=\langle \tilde g\rangle$. The following theorem formalizes this method and offers some additional properties. Algorithm~\ref{alg:lift-bch-init-from-e} specifies the process of generating and applying this manner of data encoding.

\begin{theorem}[Preservation and decoding via the binary residue]\label{thm:preserve}
Let $N$ be odd, $R=\mathbb{Z}_{2^k}[X]/(X^N+1)$, and let $\tilde g\in\mathbb{Z}_{2^k}[X]$ be the Hensel lift of a binary BCH generator $g_2\in\mathbb{F}_2[X]$ of designed distance $\delta=2t+1$ for length $N$.
Set $\mathcal{C}=\langle \tilde g\rangle\subseteq R$. Then:
\begin{enumerate}
\item $\mathcal{C}$ is an ideal of $R$, hence closed under addition and multiplication.
\item The idempotent-based encoder $Enc(m)=m\,e$ from Lemma~\ref{lem:idempotent-encoder} is a ring homomorphism with image $\mathcal{C}$.
\item Let $\pi:R\to \bar{R}:=\mathbb{F}_2[X]/(X^N-1)$ be reduction modulo $2$. Then
$\pi(\mathcal{C})=\langle g_2\rangle$ has binary Hamming distance at least $\delta$. Consequently, for any received $r=c+E$ with $c\in\mathcal{C}$ and at most $t$ nonzero symbol errors in $E$,
there is a decoder that: (i) recovers the error positions by BCH decoding of $\pi(r)$ in $\bar{R}$, and (ii) lifts the \emph{magnitudes} to modulus $2^k$ via $2$-adic Hensel lifting of the BCH key equations.
\end{enumerate}
\end{theorem}
\begin{proof} See Appendix~\S\ref{proof thm 5.3}.
\end{proof}

Note: for statement (3) part (I): more generally, recovers the error positions by BCH decoding of $\pi_u(r)$ in $\mathbb{Z}_{2^u}[X]/(X^N+1)$ for the least $u$ s.t. the syndromes are nonzero. 

Regarding parameter selecting, we choose the design parameters to expose clear cost–benefit tradeoffs and to match the pipeline’s symbol geometry. First, fix the frame length $N$ from the compute/IO batch size and FFT/NTT radix constraints. Next, pick a per-frame failure budget $\varepsilon$ (e.g., $10^{-9}$ for “nine-nines” storage/transport hops) and estimate a per-symbol SDC probability $p$ for the hop under test. Set the correction budget.
\[
t=\Big\lceil Np+\sqrt{2Np\ln(1/\varepsilon)}+\tfrac{1}{3}\ln(1/\varepsilon)\Big\rceil,\qquad \delta:=2t+1,
\]
which yields redundancy $2t$ symbols and rate $R_{\text{code}}=1-\tfrac{2t}{N}$ while keeping the dominant decoder work near $O(tN)$. For the BCH lift, select the root window $\{b,\ldots,b+\delta-2\}\pmod N$, form $g_2$ from the minimal polynomials over $\mathbb{F}_2$, and Hensel-lift to $\tilde g$; a conservative memory/compute proxy is $\deg \tilde g\le m(\delta-1)$. When erasure hints are available, favor slightly smaller $t$ and rely on the errors–erasures rule $2\tau+\rho\le 2t$ to absorb flagged symbols at unchanged overhead. If burstiness is expected, enable an automorphism interleaver $\sigma_a(X)=X^a$ with $\gcd(a,2N)=1$ and pick $a$ of large multiplicative order modulo $2N$ to maximize dispersion; this improves effective tolerance without altering the success criterion. In practice we find $t\in[4,16]$ covers $p\in[10^{-6},10^{-5}]$ at $\varepsilon=10^{-9}$ with overhead between $0.2\%$ and $1.6\%$ for the $N$ of interest, and the corresponding $\tilde O(tN)$ decoding cost remains compatible with line-rate execution on modern CPUs/GPUs, a more careful parameter tuning is shown in Section~\ref{sec:concrete-bench}.

\subsubsection{Decoding process.}
After evaluation, codewords remain in $\mathcal{C}$ by ideal closure. Upon decryption or receipt from storage/network, we correct transport faults using a two-stage decoder: first, locate errors in the binary residue where BCH is equipped; then $2$-adically lift their magnitudes to full modulus and fix the word. The following algorithm~\ref{alg:decode-lifted-bch} describes this process. Note the following: (i) The BCH step gives positions quickly and robustly; the lifting loop only performs \textbf{odd} inversions (mod $2$), avoiding divisions by even elements.  
(ii) The multiplicative encoder $m\mapsto me$ is a projector, not injective; message recovery depends on the chosen systematic section and is addressed right after this subsection.

\begin{proposition}[Injective domains]
\label{prop:inj}
Let $N$ be odd, $R=\mathbb{Z}_{2^k}[X]/(X^N+1)$, and $\mathcal{C}=Re=\langle \tilde g\rangle$ as in Theorem~\ref{thm:preserve}. Then:
\begin{enumerate}
\item The map $Enc:R\to R$, $Enc(m)=me$ is not injective for non-trivial $e$.
\item The restricted maps
\[
Enc_{\mathrm{CRT}}:\ \prod_{i\in\mathcal I} R_i \xrightarrow{\ \sim\ } Re,\qquad
Enc_{\mathrm{gen}}:\ \mathbb{Z}_{2^k}[X]/(h)\xrightarrow{\ \sim\ } \langle \tilde g\rangle
\]
Are isomorphisms. In particular, both give injective encoders.
\end{enumerate}
\end{proposition}

\begin{proof}
See Appendix~\S\ref{proof prop:inj}.
\end{proof}

\subsection{Power-of-Two Length}
\label{subsec:repeated-root}

We now consider the case when $N=2^{m}$ for some integer $m$ and $M(X)=X^{N}+1$. Then $M(X)\equiv (X+1)^{N}$ over $\mathbb{F}_2$, so $R$ is a local ring with maximal ideal $\langle 2, X+1\rangle$ and has only trivial idempotent $0,1$ as shown in Proposition~\ref{prop:no-idemp}. Thus a nontrivial ring-homomorphic encoder $Enc(m)=m e$ with $e^2=e\notin\{0,1\}$ does not exist. Nevertheless, we can enforce closure under $+$ and $\times$ inside a \emph{code ideal} and decode efficiently.

\begin{proposition}[No nontrivial idempotents when $N=2^m$]\label{prop:no-idemp}
Let $k,m\ge 1$ and $R:=\mathbb{Z}_{2^{k}}[X]/(X^{2^{m}}+1)$. Then the only idempotents
$e\in R$ (i.e., $e^2=e$) are $e=0$ and $e=1$. In particular, there is no nontrivial
ring-homomorphic encoder of the form $Enc(m)=m e$ with $e^2=e$.
\end{proposition}

\begin{proof}
See Appendix~\S\ref{proof prop no idemp}.
\end{proof}

\begin{definition}[Repeated-root negacyclic ideal codes]\label{def:RR}
For $t\ge 1$ set $\mathcal{C}_t:=\langle (X+1)^{2t}\rangle\subset R$. We claim $\mathcal{C}_t$ is an ideal of $R$.
\end{definition}

Given this ideal, we instantiate a concrete, algebra-compatible encoder and a matching bounded-distance decoder. Specifically, by Observation~\ref{obs1} the ideal $\mathcal{C}_t=\langle (X{+}1)^{2t}\rangle$ admits a \emph{systematic} map
\[
Enc:\ \{m\in R:\deg m < N-2t\}\ \longrightarrow\ \mathcal{C}_t,\qquad
Enc(m)\;=\;c(X)\;=\;m(X)\,(X{+}1)^{2t}\bmod (X^N{+}1),
\]
Which preserves addition and multiplication by arbitrary ring elements (closure in the ideal), even though Proposition~\ref{prop:no-idemp} rules out a nontrivial idempotent-based encoder in the power-of-two case. On the decoder side, Lemma~\ref{lem:hasse} supplies $2t$ \emph{Hasse-derivative syndromes} $S_i=r^{[i]}(-1)$ that annihilate all codewords, and express the received word’s deviation $r-c$ as a short linear combination of monomials with known basis functions $\big\{(-1)^{j-i}\binom{j}{i}\big\}$. These syndromes feed a standard algebraic pipeline: a short key-equation step (via Berlekamp–Massey or extended Euclid) to recover an error-locator of degree $\nu\le t$, a Chien-style search over $j\in\{0,\dots, N{-}1\}$ to find error positions, and a $\nu\times\nu$ linear solve over $\mathbb{Z}_{2^k}$ to reconstruct magnitudes; the corrected codeword $\hat c$ then yields the message by dividing out $(X{+}1)^{2t}$ in $R$. This procedure is made explicit in Algorithm~\ref{alg:decode-rr}, runs in time $\Theta(tN)$ for syndromes and locator evaluation plus $O(t^2)$ for the small solves, and extends verbatim to errors-and-erasures by replacing the first $\rho$ rows of the linear system with known erasure constraints.

\begin{observation}
\label{obs1}
(i) As a $\mathbb{Z}_{2^k}$-module, $\mathcal{C}_t$ is free of rank $N-2t$; (ii) a systematic encoder for $\mathcal{C}_t$ is $c(X)=m(X)\,(X{+}1)^{2t}\bmod (X^{N}{+}1)$ with $\deg m<N-2t$. 
\end{observation}
\begin{proof}
See Appendix~\S\ref{proof obs 1}.
\end{proof}

\begin{lemma}[Syndromes via Hasse derivatives]\label{lem:hasse}
Let $c\in \mathcal{C}_t$. Then
\(
c^{[i]}(-1)=0
\)
for all $i=0,1,\ldots,2t-1$, where $c^{[i]}$ denotes the $i$-th Hasse derivative.
Moreover, for any received word $r=c+E$ with $E=\sum_{\ell=1}^{\nu} e_\ell X^{j_\ell}$ and $\nu\le t$, the syndromes
\[
S_i\ :=\ r^{[i]}(-1)\ =\ \sum_{\ell=1}^{\nu} e_\ell\,(-1)^{\,j_\ell-i}\binom{j_\ell}{i}\qquad (i=0,1,\dots,2t-1).
\]
They are given by the above closed form.
\end{lemma}

\begin{proof}
See Appendix~\S\ref{proof lem hasse}.
\end{proof}

\subsubsection{Single-error closed form.}
For the repeated-root code with $t=1$, a received word of the form $r=c+eX^{j}$ has
Hasse-derivative syndromes (Lemma~\ref{lem:hasse})
\[
S_0=e(-1)^j,\qquad S_1=j\,e(-1)^{j-1},\qquad S_2=\binom{j}{2}\,e(-1)^{j-2}.
\]
These are just the first three “moments’’ of the single spike at position $j$ with
magnitude $e$. Eliminating $e$ via the invariant $D\:=\ S_1^2-2S_0S_2\ =\ S_0^2\,j$ where we need to avoid dividing by a potentially even $e$. Write $S_0=2^{v}s_0$ with $s_0$ odd
($v=\nu_2(e)$), so $S_0^2=2^{2v}s_0^2$, and recover the location by
\[
j \ \equiv\ \Big(\tfrac{D}{2^{2v}}\Big)\,s_0^{-2}\ \ (\bmod\ 2^{\,k-2v}).
\]
When $2^{k-2v}\ge N$, this identifies $j\in\{0,\dots,N-1\}$ uniquely; if not,
one can disambiguate using $S_3$ (or by checking candidates against $S_0,S_1,S_2$).
Finally, the magnitude follows without division by even numbers: $e \ \equiv\ S_0\,(-1)^{j}\ \ (\bmod\ 2^{k})$, where $(-1)^{j}\in\{\pm 1\}$ is determined by the parity of the recovered $j$.
This constant-time closed form is robust even when $e$ is highly even (large $v$),
because we normalize out $2^{2v}$ before inverting the odd unit $s_0^2$.

\subsubsection{Error-correction Capability and Parametrization.}
The above decoder is bounded-distance: it uniquely corrects any $\tau\le t$ symbol errors,
and, with a short CRC, it detects patterns beyond that radius. The closed-form we show
previously is only for the single-error case $t{=}1$; for $\tau\ge2$ we use the
general Hasse-syndrome $\to$ locator (BM/EEA) $\to$ magnitude-lifting procedure. The
redundancy is exactly $2t$ symbols, so the rate is $R=1-\tfrac{2t}{N}$ (requiring $2t<N$).
If $\rho$ symbol locations are marked as erasures, decoding succeeds whenever
$2\tau+\rho\le 2t$. However, we do have an interest in parameterizing $t$. Assume i.i.d.\ symbol errors with probability $p$ per coefficient and target per-frame
failure budget $\varepsilon$. Let $X\sim\mathrm{Bin}(N,p)$ count symbol errors; we want $\Pr[X>t]\ \le\ \varepsilon$. We have the Chernoff sizing, which offers a rigorous upper bound as follows. For any $\tau>0$,
\[
\Pr[X\ge \mu+\tau]\ \le\ \exp\!\Big(-\,\frac{\tau^{2}}{2(\mu+\tau/3)}\Big).
\]
Thus, it suffices to pick $\tau$ so that
$\frac{\tau^{2}}{2(\mu+\tau/3)}\ge \ln(1/\varepsilon)$; a convenient closed form is
\[
\tau\ =\ \sqrt{\,2\mu\ln(1/\varepsilon)\,}\;+\;\tfrac{1}{3}\ln(1/\varepsilon),
\qquad
t\;=\;\Big\lceil\,\mu+\tau\,\Big\rceil,
\]
which guarantees $\Pr[X>t]\le\varepsilon$.

\subsection{Automorphism-Based Interleaving for Burst-Resilience}
\label{sec:automorphism-interleave}

\begin{figure}[t]
  \centering
  \includegraphics[width=1.0\linewidth,trim=0cm 15.5cm 0cm 1cm,clip]{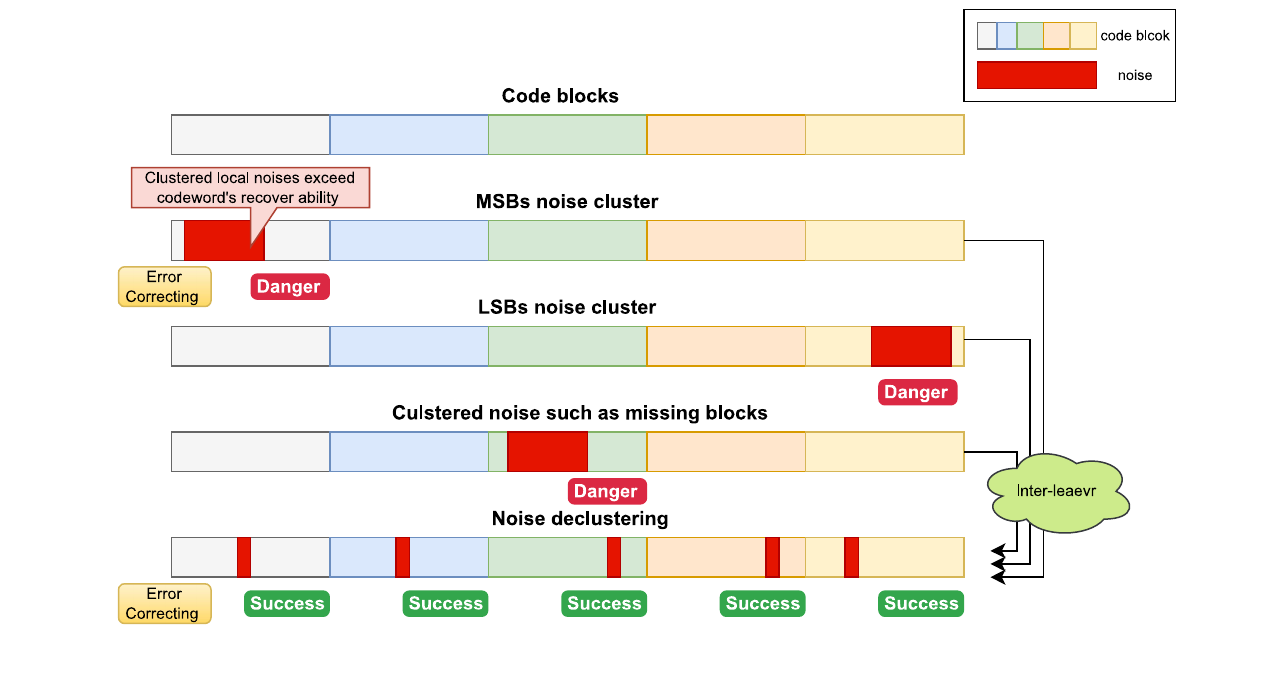}
  \caption{Logistics of interleaver}
  \label{fig:intlv}
\end{figure}

In practice, code length and rate constraints require segmenting data into shorter frames, and transport errors often exhibit locality. We propose an
automorphism-based interleaver that (i) permutes polynomial coordinates
via a ring automorphism to disperse clustered errors, and (ii) preserves the
algebra needed by HE evaluation and our codes. The functionality of our interleaver is illustrated in Figure~\ref{fig:intlv}

\begin{lemma}\label{lem:sigma-automorphism}
Let $R=\mathbb{Z}_{2^k}[X]/(X^N+1)$ and let $a$ be an odd integer with $\gcd(a,2N)=1$.
Define $\sigma_a:R\to R$ on representatives by
\[
\sigma_a\ \!(\sum_{j=0}^{N-1} u_j X^j)\ :=\ \sum_{j=0}^{N-1} u_j\,X^{aj}\ \ (\bmod\ X^N+1),
\]
i.e., $\sigma_a(X)=X^a$ and $\sigma_a$ fixes coefficients $u_j\in\mathbb{Z}_{2^k}$.
Then $\sigma_a$ is a ring automorphism of $R$.
\end{lemma}

\begin{proof}
See Appendix~\S\ref{proof lem 5.8}.
\end{proof}

\begin{proposition}[Preservation of HE semantics]\label{prop:commute}
Let $f$ be any circuit over $R$ built from additions, multiplications, and constants in $\mathbb{Z}_{2^k}$.
For all inputs $m_1,\dots,m_r\in R$ and any odd $a$ with $\gcd(a,2N)=1$,
\[
\sigma_a\big(f(m_1,\dots,m_r)\big)\;=\;f\big(\sigma_a(m_1),\dots,\sigma_a(m_r)\big).
\]
\end{proposition}

\begin{proof}
See Appendix~\S\ref{subsec:app_commute}.
\end{proof}

\section{Benchmark}
\label{sec:concrete-bench}
In this section, we give concrete, implementation-oriented codes for four power-of-two frame lengths of 1024, 2048, 4096, 8192 and for their neighboring odd lengths 1025, 2049, 4097, 8193. For power-of-two sizes, we use a repeated-root encoder implemented as a single negacyclic ring multiplication by a precomputed parity mask; for the odd sizes, we use an idempotent projector, also realized by one ring multiplication. Both families add exactly twice the design parameter $t$ in parity symbols, so the rate loss is small and predictable. Decoding succeeds whenever the weighted sum of impairments—counting each unknown symbol error twice and each flagged erasure once—does not exceed the budget determined by $t$. Using a standard Chernoff sizing under independent symbol errors with a per-symbol error probability of $10^{-6}$ (Setting A) and a per-frame failure target of $10^{-9}$, we obtain $t=8$ across all lengths; under a more stressed $10^{-5}$ (Setting B) we still have $t=8$ except $t=9$ at 4096/8192 and their odd counterparts. The resulting overheads range from about $1.56\%$ at 1024 down to about $0.20\%$ at 8192, with essentially identical numbers for the odd lengths, yielding effective code rates of at least 0.984. With a simple coefficient-permutation interleaver, contiguous byte bursts are dispersed so that, for 32-bit and 64-bit symbols, the system tolerates unknown-error bursts up to roughly 32/64~bytes for $t=8$ (and 36/72~bytes for $t=9$), or twice those sizes when the bytes are flagged as erasures. Encoding cost is one ring multiply per frame; decoding cost scales linearly with $t$ and the frame length for syndrome computation plus a small, $t$-sized algebraic solve. For the odd-length encoders, modest field-size parameters imply generator-polynomial degrees bounded by a few hundred coefficients, which are small compared to the frame length and straightforward to provision.

\noindent\textbf{Notation}
We work in $R=\mathbb{Z}_{2^k}[X]/(X^N+1)$ with $N\in\{1024,2048,4096,8192\}$ and the induced odd lengths $\{1025,2049,4097,8193\}$.
A frame is a polynomial $m(X)=\sum_{j=0}^{N-1} m_j X^j\in R$ whose coefficients are $k$-bit words.
For an integer $t\ge1$, both constructions add exactly $2t$ parity symbols, yielding rate
$R_{\text{code}}=1-\tfrac{2t}{N}$ and minimum distance $d_{\min}\ge 2t+1$.
A mixture of unknown symbol errors $\tau$ and erasures $\rho$ is correctable whenever $2\tau+\rho\le 2t$.

\subsection{Power-of-two Case}
Define the fixed parity mask
\[
h_{N,t}(X)\;\equiv\;(X+1)^{2t}\bmod (X^N+1)\in R.
\]
The systematic codeword is the negacyclic product.
\[
c(X)\;=\;m(X)\,h_{N,t}(X)\;\bmod (X^N+1).
\]
Encoding costs one ring multiply per frame (FFT/NTT-based $O(N\log N)$).
A practical decoder computes $2t$ Hasse-derivative syndromes at $X=-1$ ($\approx 2t\, N$ word operations),
solves a small locator/magnitude system ($O(t^2)$), then applies a sparse correction.

\subsection{Odd Case}
Let $g_{N,t}(X)\in\mathbb{Z}_{2^k}[X]$ be a Hensel lift of a binary BCH generator of designed distance $\delta=2t+1$
for length $N$; equivalently, let $e_{N,t}\in R$ be the associated CRT idempotent projector.
Either form of the encoder preserves ring operations:
\[
c(X)\;=\;m(X)\,g_{N,t}(X)\;\bmod (X^N+1)\;\;\;\text{or}\;\;\;c(X)\;=\;m(X)\,e_{N,t}.
\]
Decoding uses a binary BCH locator, e.g., BM/EEA+Chien; $\tilde O(tN)$ over $\mathbb{F}_{2^{m_N}}$ with $m_N=\mathrm{ord}_N(2)$,
followed by $(k-1)$ $2$-adic magnitude lifts ($O(t^2)$ total).

\subsection{Benchmark Result}
Given per-symbol error probability $p$ and per-frame failure budget $\varepsilon$,
size
\[
t \;=\;\Big\lceil Np + \sqrt{2Np\ln(1/\varepsilon)} + \tfrac{1}{3}\ln(1/\varepsilon) \Big\rceil
\]
so that $\Pr[X>t]\le \varepsilon$ for $X\sim\mathrm{Bin}(N,p)$.
We instantiate two practical regimes: \emph{Setting A} ($p=10^{-6}$) and \emph{Setting B} ($p=10^{-5}$), both with $\varepsilon=10^{-9}$. Tables~\ref{tab:pow2} and \ref{tab:odd} instantiate these settings across all lengths, reporting the chosen $t$ and the resulting rate/overhead.

\begin{table}[t]
\centering
\begin{minipage}[t]{0.485\linewidth}
\centering
\scriptsize
\captionof{table}{Power-of-two lengths. Parameters under Setting A ($p=10^{-6}$) and Setting B ($p=10^{-5}$).}
\label{tab:pow2}
\begingroup\setlength{\tabcolsep}{4pt}
\resizebox{\linewidth}{!}{%
\begin{tabular}{rcccccc}
\toprule
& \multicolumn{3}{c}{Setting A} & \multicolumn{3}{c}{Setting B} \\
\cmidrule(lr){2-4}\cmidrule(lr){5-7}
$N$ & $t$ & Ovhd $2t/N$ & Rate & $t$ & Ovhd $2t/N$ & Rate \\
\midrule
1024 & 8 & 1.562\% & 0.984375 & 8 & 1.562\% & 0.984375 \\
2048 & 8 & 0.781\% & 0.992188 & 8 & 0.781\% & 0.992188 \\
4096 & 8 & 0.391\% & 0.996094 & 9 & 0.439\% & 0.995605 \\
8192 & 8 & 0.195\% & 0.998047 & 9 & 0.220\% & 0.997803 \\
\bottomrule
\end{tabular}}
\endgroup
\par\medskip\footnotesize Ovhd = overhead; Rate $=R_{\text{code}}$.
\end{minipage}\hfill
\begin{minipage}[t]{0.485\linewidth}
\centering
\scriptsize
\captionof{table}{Induced odd lengths (lifted BCH / idempotent). Same settings as Table~\ref{tab:pow2}.}
\label{tab:odd}
\begingroup\setlength{\tabcolsep}{4pt}
\resizebox{\linewidth}{!}{%
\begin{tabular}{rcccccc}
\toprule
& \multicolumn{3}{c}{Setting A} & \multicolumn{3}{c}{Setting B} \\
\cmidrule(lr){2-4}\cmidrule(lr){5-7}
$N$ & $t$ & Ovhd $2t/N$ & Rate & $t$ & Ovhd $2t/N$ & Rate \\
\midrule
1025 & 8 & 1.561\% & 0.984390 & 8 & 1.561\% & 0.984390 \\
2049 & 8 & 0.781\% & 0.992191 & 8 & 0.781\% & 0.992191 \\
4097 & 8 & 0.391\% & 0.996095 & 9 & 0.439\% & 0.995607 \\
8193 & 8 & 0.195\% & 0.998047 & 9 & 0.220\% & 0.997803 \\
\bottomrule
\end{tabular}}
\endgroup
\par\medskip\footnotesize Ovhd = overhead; Rate $=R_{\text{code}}$.
\end{minipage}
\end{table}
For a chosen $(N,t)$, the extremal budgets are $\tau_{\max}=t$ and $\rho_{\max}=2t$.
The slack above the mean error count is $t-Np$; under Setting~A this slack is $\approx 8- N\cdot 10^{-6}$,
and under Setting~B it is $7.98$–$8.959$ depending on $N$, the following Table~\ref{tab:bursts} shows the computation results. 

Regarding complexity proxies per frame, let $M(N)$ denote one negacyclic ring multiply via FFT/NTT with cost $\Theta(N\log N)$. Encoding uses $1\times M(N)$ in both families. A repeated-root decoder performs $2t$ syndromes $\approx 2t\,N$ word ops plus $O(t^2)$ algebra;
a BCH decoder performs $2t$ binary syndromes, one Chien sweep , requiring about $N$ test points with a degree-$\le t$ locator,
and $(k-1)$ $2$-adic lifts with $O(t^2)$ total. For the concrete grid where word-operation counts are shown for the syndrome phase, we have the following Table~\ref{tab:complexity}. For reference, the FFT size factors $N\log_2 N$ across these lengths are
$10{,}240$, $22{,}528$, $49{,}152$, $106{,}496$ (power-of-two) and
$10{,}251$, $22{,}540$, $49{,}165$, $106{,}510$ (odd), respectively.

\begin{table}[t]
\centering
\begin{minipage}[t]{0.485\linewidth}
\centering
\scriptsize
\captionof{table}{Burst and erasure tolerance per frame (bytes) after interleaving.
Symbol size $s\in\{4,8\}$ bytes corresponds to $k\in\{32,64\}$.}
\label{tab:bursts}
\begingroup\setlength{\tabcolsep}{3.5pt}
\resizebox{\linewidth}{!}{%
\begin{tabular}{rcccccccc}
\toprule
& \multicolumn{4}{c}{Setting A ($p=10^{-6}$; $t=8$)} & \multicolumn{4}{c}{Setting B ($p=10^{-5}$; $t$ as shown)} \\
\cmidrule(lr){2-5}\cmidrule(lr){6-9}
$N$ & $s$ & $B_{\text{unk,max}}$ & $B_{\text{era,max}}$ & & $s$ & $t$ & $B_{\text{unk,max}}$ & $B_{\text{era,max}}$ \\
\midrule
1024 & 4/8 & 32/64  & 64/128 & & 4/8 & 8  & 32/64  & 64/128 \\
2048 & 4/8 & 32/64  & 64/128 & & 4/8 & 8  & 32/64  & 64/128 \\
4096 & 4/8 & 32/64  & 64/128 & & 4/8 & 9  & 36/72  & 72/144 \\
8192 & 4/8 & 32/64  & 64/128 & & 4/8 & 9  & 36/72  & 72/144 \\
1025 & 4/8 & 32/64  & 64/128 & & 4/8 & 8  & 32/64  & 64/128 \\
2049 & 4/8 & 32/64  & 64/128 & & 4/8 & 8  & 32/64  & 64/128 \\
4097 & 4/8 & 32/64  & 64/128 & & 4/8 & 9  & 36/72  & 72/144 \\
8193 & 4/8 & 32/64  & 64/128 & & 4/8 & 9  & 36/72  & 72/144 \\
\bottomrule
\end{tabular}}
\endgroup
\end{minipage}\hfill
\begin{minipage}[t]{0.485\linewidth}
\centering
\scriptsize
\captionof{table}{Decoder syndrome-operation counts per frame (dominant term).}
\label{tab:complexity}
\begingroup\setlength{\tabcolsep}{4pt}
\resizebox{\linewidth}{!}{%
\begin{tabular}{rcccc}
\toprule
$N$ & $2t$ (A) & Syndrome ops (A) & $2t$ (B) & Syndrome ops (B) \\
\midrule
1024 & 16 & 16{,}384  & 16 & 16{,}384  \\
2048 & 16 & 32{,}768  & 16 & 32{,}768  \\
4096 & 16 & 65{,}536  & 18 & 73{,}728  \\
8192 & 16 & 131{,}072 & 18 & 147{,}456 \\
1025 & 16 & 16{,}400  & 16 & 16{,}400  \\
2049 & 16 & 32{,}784  & 16 & 32{,}784  \\
4097 & 16 & 65{,}552  & 18 & 73{,}746  \\
8193 & 16 & 131{,}088 & 18 & 147{,}474 \\
\bottomrule
\end{tabular}}
\endgroup
\end{minipage}
\end{table}

Now we give algebraic parameters for Odd-length cases. Let $m_N=\mathrm{ord}_N(2)$ denote the multiplicative order of $2$ modulo $N$, therefore a primitive $N$-th root of unity lies in $\mathbb{F}_{2^{m_N}}$.
For designed distance $\delta=2t+1$, a standard upper bound is $\deg g_{N,t}\le (\delta-1)\,m_N$. The values on our grid are shown in Table~\ref{tab:odd-length-params}.  These bounds guide memory/compute provisioning when implementing the generator-polynomial path; the idempotent projector $e_{N,t}$ is an equivalent encoder.

\begin{table}[t]
\centering
\small
\caption{Odd-length parameters. Let $m_N=\operatorname{ord}_N(2)$ so a primitive $N$-th root of unity lies in $\mathbb{F}_{2^{m_N}}$. For designed distance $\delta=2t+1$, a standard bound is $\deg g_{N,t}\le (\delta-1)m_N$.}
\label{tab:odd-length-params}
\begin{tabular}{rcccc}
\toprule
$N$ & $m_N=\operatorname{ord}_N(2)$ & $\deg g_{N,8}\le 16\,m_N$ & $\deg g_{N,9}\le 18\,m_N$ & $\deg g_{N,8}/N$ \\
\midrule
1025 & 20 & $\le 320$ & $\le 360$ & $\le 0.312$ \\
2049 & 22 & $\le 352$ & $\le 396$ & $\le 0.172$ \\
4097 & 24 & $\le 384$ & $\le 432$ & $\le 0.094$ \\
8193 & 26 & $\le 416$ & $\le 468$ & $\le 0.051$ \\
\bottomrule
\end{tabular}
\end{table}

\section{Conclusion and Future Work}
We set out to protect polynomially encoded frames as they move across different computation stages, where rare but costly silent corruptions can derail downstream computation~\citep{Schroeder2009DRAMWild}. Our approach is a ring-compatible reliability layer that lives in the same algebra as the data, adds systematic redundancy, and corrects symbol errors as well as flagged erasures without format conversions or round trips. We instantiated this layer with two complementary codes that cover common frame lengths, a repeated-root negacyclic design for powers of two, and a Hensel-lifted BCH design with an idempotent encoder for odd lengths, and equipped them with an automorphism-based interleaver that disperses bursty faults while preserving code membership. These choices keep protection on the fast path: redundancy is precise, encode/decode costs scale with frame size, and code membership is preserved through linear stages and multiplicative stages as well for ideal-based encoders inspired by former works in algebraic coding theory~\citep{MacWilliamsSloane1977}. The layer also composes cleanly with CRCs by treating their flags as erasures, expanding the correctable region without extra metadata.

Looking ahead, several directions can deepen the impact and broaden applicability. First, adaptive tuning that uses online error telemetry to pick the correction budget and interleaver on the fly would align overheads with observed SDC rates and burst profiles~\citep{Schroeder2009DRAMWild,Meza2015SIGMETRICS,Bairavasundaram2008FAST}. Second, hardware offload on NICs, DPUs, and GPUs—alongside kernel-bypass I/O paths—can drive latency down while sustaining line-rate throughput~\citep{DPDKGuide,NvidiaBluefield,PCISIG2021PCIe6}. Third, extending the encoder/decoder toolkit to mixed-modulus pipelines and to multi-frame streaming interleavers would cover a wider range of analytics and ML workloads~\citep{HuffmanPless2003,LinCostello2004}. Fourth, tighter finite-length analyses for miscorrection probability and heavy-tailed bursts, plus end-to-end evaluations under realistic traffic mixes, would give operators crisp SLO-to-parameter mappings~\citep{Bossert2021BCH}. Finally, co-design with algorithm-based fault tolerance and coded computation can yield complementary protection across execution and transport~\citep{Huang1984ABFT,Yu2017PolynomialCodes}.

\bibliographystyle{ACM-Reference-Format}
\bibliography{refs}
\appendix
\newpage
\section{Algorithms}

\noindent\textbf{\textsc{InvModHensel}$(a,m,k)$.}
Algorithm~\ref{alg:invmod-hensel} computes the inverse of $a$ modulo a monic $m$ over $\mathbb{Z}_{2^k}$ by first inverting $\bar a$ mod $\bar m$ in $\mathbb{F}_2[X]$ and then Hensel-lifting the inverse coefficientwise from modulus $2$ up to $2^k$, correcting at each lift step so that $a\,u\equiv 1 \pmod{m}$ holds with coefficients in $\mathbb{Z}_{2^k}$.

\begin{algorithm}[H]
\small
\caption{\textsc{InvModHensel}$(a,m,k)$ — inverse of $a$ modulo $m$ over $\mathbb{Z}_{2^k}$}
\label{alg:invmod-hensel}
\begin{algorithmic}[1]
\Require $a,m\in\mathbb{Z}_{2^k}[X]$, $m$ monic, $\gcd(\bar a,\bar m)=1$ in $\mathbb{F}_2[X]$; integer $k\ge 1$
\Ensure $u\in \mathbb{Z}_{2^k}[X]/(m)$ with $a\,u\equiv 1 \pmod{m}$ (coefficients mod $2^k$)
\State Compute $\bar u \gets a^{-1}\!\!\pmod{\bar m}$ via EEA in $\mathbb{F}_2[X]$
\State Lift $\bar u$ to $u^{(1)}$ (same 0/1 coeffs) in $\mathbb{Z}_{2}[X]/(m)$
\For{$t=1$ \textbf{to} $k-1$} \Comment{Hensel lift: from $2^{t}$ to $2^{t+1}$}
  \State $r \gets 1 - a\,u^{(t)} \pmod{(m,\,2^{t+1})}$ \hfill \small{all polys mod $m$, coeffs mod $2^{t+1}$}
  \If{$r \equiv 0 \ (\mathrm{mod}\ 2^{t+1})$} \State $u^{(t+1)} \gets u^{(t)}$ \textbf{continue} \EndIf
  \State $\rho \gets (r/2^{t}) \bmod 2 \ \in \mathbb{F}_2[X]/(\bar m)$
  \State $w \gets \bar a^{-1}\cdot \rho \ \bmod \bar m$ \hfill \small{work in $\mathbb{F}_2[X]$}
  \State Lift $w$ to $\widehat w$ with coeffs in $\{0,1\}\subset\mathbb{Z}_{2^{t+1}}$
  \State $u^{(t+1)} \gets u^{(t)} + 2^{t}\,\widehat w \ \ (\bmod\ m,\,2^{t+1})$
\EndFor
\State \textbf{return} $u^{(k)}$
\end{algorithmic}
\end{algorithm}

\noindent\textbf{\textsc{DecodeLiftedBCH}$(r;\,t,k,\delta,b)$.}
Algorithm~\ref{alg:decode-lifted-bch} decodes a lifted BCH codeword in $R=\mathbb{Z}_{2^k}[X]/(X^N{+}1)$ by reducing $r$ mod $2$ to locate error \emph{positions} via standard binary BCH, then iteratively lifting error \emph{magnitudes} from $\{0,1\}$ to $\mathbb{Z}_{2^k}$ using small linear updates modulo $2,4,\dots,2^k$; finally subtracts the lifted error pattern and optionally recovers the message from the systematic form.

\begin{algorithm}[H]
\small
\caption{\textsc{DecodeLiftedBCH}$(r;\,t,k,\delta,b)$}
\label{alg:decode-lifted-bch}
\begin{algorithmic}[1]
\Require Received word $r\in R=\mathbb{Z}_{2^k}[X]/(X^N{+}1)$; error capability $t$; modulus power $k$; BCH parameters $(\delta{=}2t{+}1,\ b)$; optional erasures from CRC
\Ensure Corrected codeword $\hat c\in\mathcal{C}$ (and application message if a systematic map is in place)
\State \textbf{Binary residue.} Reduce coefficients mod $2$ to get $r_2\in\mathbb{F}_2[X]/(X^N{+}1)$
\State \textbf{Locate errors.} Run standard binary BCH decoding on $r_2$ (with any erasures) to identify up to $t$ error \emph{positions} $J=\{j_1,\dots,j_\nu\}$; record binary magnitudes $a_\ell\in\{0,1\}$
\State \textbf{Initialize magnitudes.} Set provisional error values $e_\ell^{(1)}\gets a_\ell$ for all $j_\ell\in J$
\For{$u=1$ \textbf{to} $k{-}1$} 
  \State Recompute a small set of \emph{syndromes} of $r$ modulo $2^{u+1}$ (e.g., the first $2t$ BCH/Hasse syndromes)
  \State Form the residual “what’s still wrong at this modulus” using the current guess $e^{(2^u)}=\{e_\ell^{(2^u)}\}$
  \State \textbf{Solve a tiny linear system over $\mathbb{F}_2$} (size $\le 2t$): obtain $\Delta e\in\{0,1\}^{\nu}$ indicating which $e_\ell$ need a $+2^u$ bump
  \State Update magnitudes: $e_\ell^{(2^{u+1})}\gets e_\ell^{(2^u)} + 2^u\cdot \Delta e_\ell$
\EndFor
\State \textbf{Correct the word.} Subtract the lifted error pattern:
       $\hat c(X)\gets r(X)-\sum_{\ell=1}^{\nu} e_\ell^{(2^k)} X^{j_\ell}$ \hfill \small{now $\hat c\in\mathcal{C}$}
\State \textbf{Recover application message.} If a \emph{systematic} mapping $R\!\to\!\mathcal{C}$ was used, return $\hat m$ by dividing out the generator in $R$. If encoding was by the projector $m\mapsto me$, return $\hat c$ or apply the agreed section $Re\!\to\!\mathcal{M}$ (see next subsection).
\State \textbf{return} $\hat c$ and $\hat m$
\end{algorithmic}
\end{algorithm}

\noindent\textbf{\textsc{DecodeRR}$(r;\,t,k)$.}
Algorithm~\ref{alg:decode-rr} decodes the repeated-root code $\langle (X{+}1)^{2t}\rangle$ (for $N=2^m$) using Hasse-derivative syndromes at $X=-1$: if all syndromes vanish, divide by $(X{+}1)^{2t}$; otherwise solve the key equation (BM/EEA) to obtain the error locator, find error positions (Chien-like search), solve a small linear system for magnitudes over $\mathbb{Z}_{2^k}$, correct $r$, and divide to recover the message.

\begin{algorithm}[t]
\small
\caption{\textsc{DecodeRR}$(r;\,t,k)$ \quad (Repeated-root, $N=2^{m}$, no interleaving)}
\label{alg:decode-rr}
\begin{algorithmic}[1]
\Require Received word $r\in R=\mathbb{Z}_{2^{k}}[X]/(X^{2^{m}}{+}1)$, error capability $t\ge 1$
\Ensure Corrected codeword $\hat c\in \mathcal C_t=\langle (X{+}1)^{2t}\rangle$ and message $\hat m$
\State \textbf{Syndromes (Lemma~\ref{lem:hasse}):} $S_i \gets r^{[i]}(-1)$ for $i=0,\dots,2t{-}1$
\If{$S_i=0$ for all $i$} \label{line:rr-clean}
  \State $\hat c \gets r$;\quad $\hat m \gets \hat c/(X{+}1)^{2t}$ in $R$;\quad \textbf{return} $(\hat c,\hat m)$
\EndIf
\State \textbf{Key equation:} from $(S_i)$ compute locator $\Lambda(z)$ and evaluator $\Omega(z)$
       via Berlekamp--Massey or EEA (degree $\nu\!\le\! t$)
\State \textbf{Positions:} find $J=\{j_1,\ldots,j_\nu\}$ by a Chien-like search for roots of $\Lambda$
\State \textbf{Magnitudes:} solve the $\nu\times \nu$ linear system
       $\sum_{\ell=1}^{\nu} e_\ell\,(-1)^{\,j_\ell-i}\binom{j_\ell}{i}=S_i$ for $i=0,\dots,\nu{-}1$
       over $\mathbb{Z}_{2^{k}}$ to obtain $e_\ell$
\State \textbf{Correction:} $\hat c \gets r - \sum_{\ell=1}^{\nu} e_\ell X^{j_\ell}$
\State \textbf{Message (systematic):} $\hat m \gets \hat c/(X{+}1)^{2t}$ in $R$
\State \textbf{return} $(\hat c,\hat m)$
\end{algorithmic}
\end{algorithm}

\noindent\textbf{\textsc{LiftBCHEncoderInitFromIdempotents}$(\cdot)$.}
Algorithm~\ref{alg:lift-bch-init-from-e} initializes a lifted BCH encoder for odd $N$ by selecting cyclotomic exponent sets $\mathcal S=\{b,\dots,b+\delta-2\}$, aggregating the corresponding primitive idempotents $e_i$ to form the projector $e=\sum_{i\in\mathcal I}e_i$, and (optionally) the polynomial generator $\tilde g=\prod_{i\in\mathcal I}\tilde f_i$; returns state tying $Re=\langle \tilde g\rangle$ to $(\delta,b)$ for encoding/decoding.

\begin{algorithm}[t]
\small
\caption{\textsc{LiftBCHEncoderInitFromIdempotents}$(N,k,\delta,b;\ \{f_i\},\{\tilde f_i\},\{e_i\})$}
\label{alg:lift-bch-init-from-e}
\begin{algorithmic}[1]
\Require Odd $N$, modulus power $k\!\ge\!1$, designed distance $\delta=2t{+}1$, start exponent $b$;
         binary factors $\{f_i\}$ of $X^N\!+\!1$ over $\mathbb F_2$, their Hensel lifts $\{\tilde f_i\}$ in $\mathbb Z_{2^k}[X]$, primitive idempotents $\{e_i\}$ in $R=\mathbb Z_{2^k}[X]/(X^N{+}1)$
\Ensure State $\mathsf{st}=(e,\tilde g,\mathcal I,\delta,b)$ with code $\mathcal C=Re=\langle\tilde g\rangle$
\State $\mathcal S \gets \{\,b,\,b{+}1,\,\dots,\,b{+}\delta{-}2\,\}\ (\bmod N)$
\State $\mathcal I \gets \{\, i : \text{$f_i$’s cyclotomic exponent set }C_i \subseteq \mathcal S\,\}$
\State $e \gets \sum_{i\in\mathcal I} e_i$
\State $\tilde g(X)\gets \prod_{i\in\mathcal I}\tilde f_i(X)$ \Comment{optional, for polynomial form}
\State \textbf{return } $\mathsf{st}=(e,\tilde g,\mathcal I,\delta,b)$
\end{algorithmic}
\end{algorithm}

\section{Proofs}
\subsection{Proof for lemma~\ref{lem:idempotent-encoder}}
\label{proof lem enc}
\begin{proof}
Because the $\tilde f_i$ are pairwise coprime, the Chinese Remainder Theorem yields a ring isomorphism
\[
\Phi:\ R=\mathbb{Z}_{2^k}[X]/(M) \;\xrightarrow{\ \sim\ }\; \prod_{i=1}^s R_i,\qquad
R_i:=\mathbb{Z}_{2^k}[X]/(\tilde f_i),
\]
given by reduction modulo each $\tilde f_i$.
Let $\pi_i:R\to R_i$ denote the $i$th component of $\Phi$.

By construction, for each $i$,
\[
e_i \equiv 1 \pmod{\tilde f_i}\quad\text{and}\quad e_i \equiv 0 \pmod{\tilde f_j}\ \ (j\neq i),
\]
because $M_i$ vanishes modulo $\tilde f_j$ for $j\neq i$ and $u_i$ is an inverse of $M_i$ modulo $\tilde f_i$.
Hence in the product ring one has
\[
\Phi(e_i)=(0,\dots,0,\underbrace{1}_{i\text{th}},0,\dots,0).
\]
It follows immediately that
$e_i^2=e_i$, $e_ie_j=0$ for $i\neq j$, and $\sum_{i=1}^s e_i=1$ in $R$ (componentwise identities in the product).
For $e=\sum_{i\in\mathcal I} e_i$ we then get $e^2=\sum_{i\in\mathcal I} e_i^2+\sum_{i\neq j\in\mathcal I} e_i e_j=e$, proving (1).

Now we want to show $Enc$ map is indeed a Ring homomorphism. Additivity of $Enc$ is clear. For multiplicativity, we have 
\[
Enc(ab)=(ab)e=(ab)e^2=(ae)(be)=Enc(a)Enc(b),
\]
using $e^2=e$ and abelian of $R$.
By definition of $Enc$, $\operatorname{Im}(Enc)=\{me:m\in R\}=Re=\mathcal C$, proving (2).

Let $\bar{\mathcal I}$ be the complement of $\mathcal I$ and set $h:=h_{\bar{\mathcal I}}=\prod_{j\in\bar{\mathcal I}} \tilde f_j$. Consider the ideal $\langle h\rangle= hR\subseteq R$. We compare its CRT image with that of $Re$.

First, $\Phi(Re)$ consists of all tuples whose $i$th component is arbitrary for $i\in\mathcal I$
and zero for $i\notin\mathcal I$, because $\Phi(e)$ has $i$th component $1$ for $i\in\mathcal I$
and $0$ otherwise, and multiplication by $e$ acts as the projection onto those coordinates.

Second, reduce $h$ modulo each $\tilde f_i$:
\[
h \equiv
\begin{cases}
0 \pmod{\tilde f_i}, & i\in \bar{\mathcal I},\\
\text{unit in }R_i, & i\in \mathcal I,
\end{cases}
\]
since $h$ contains the factor $\tilde f_i$ exactly when $i\in\bar{\mathcal I}$, and is coprime to $\tilde f_i$ otherwise.
Therefore,
\[
\Phi(\langle h\rangle) \;=\; \prod_{i=1}^s \langle h \bmod \tilde f_i\rangle
\;=\; \Big(\prod_{i\in\mathcal I} R_i\Big)\ \times\ \Big(\prod_{i\in\bar{\mathcal I}} \{0\}\Big),
\]
which is the same subset of the product ring as $\Phi(Re)$.
Since $\Phi$ is an isomorphism, we conclude $\langle h\rangle = Re$, proving (3).
\end{proof}

\subsection{Proof for Theorem~\ref{thm:preserve}}
\label{proof thm 5.3}
\begin{proof}
(1)–(2) are from above lemma. For (3): Let $\pi:R\to\bar R:=\mathbb{F}_2[X]/(X^N-1)$ be reduction modulo $2$.
Because $\tilde g\bmod 2=g_2$, we have
$\pi(\mathcal{C})=\langle g_2\rangle\subseteq\bar R$, and it
has binary Hamming distance at least $\delta=2t+1$ by BCH bound. It remains to show that any $r=c+E$ with $c\in\mathcal{C}$ and at most $t$ nonzero
symbol errors in $E=\sum_{\ell=1}^{\nu} e_\ell X^{j_\ell}$ ($\nu\le t$) can be decoded
to $c$ by: (i) identifying the error positions $\{j_\ell\}$ via a BCH step on a suitable
binary residue, and (ii) recovering the magnitudes $e_\ell\in\mathbb{Z}_{2^k}$ by
$2$-adic lifting. We proceed in four steps.

\smallskip
\noindent\textbf{(A) Binary residue distance and the BCH syndromes.}
Fix a primitive $N$-th root of unity $\alpha\in\mathbb{F}_{2^s}$
(where $s=\mathrm{ord}_N(2)$), so that the binary BCH code
$\bar{\mathcal{C}}=\langle g_2\rangle$ is defined by the
zero-loci $\{\alpha^b,\alpha^{b+1},\dots,\alpha^{b+\delta-2}\}$ in the usual way.
Let $\tilde\alpha$ be the Hensel lift of $\alpha$ to an unramified extension
$\mathbb{Z}_{2^k}[\tilde\alpha]$ (i.e., $\tilde f(\tilde\alpha)=0$ where
$\tilde f$ is the lift of the minimal polynomial of $\alpha$).
For $h=0,1,\dots,\delta-2$ define the \emph{(BCH) syndromes}
\[
S_h \ :=\ r(\tilde\alpha^{\,b+h})
\ =\ \sum_{\ell=1}^{\nu} e_\ell\, \tilde\alpha^{(b+h)j_\ell}\ \ \in\ \mathbb{Z}_{2^k}[\tilde\alpha].
\]
Because $c\in\mathcal{C}$ vanishes on these evaluation points, $S_h$ depends only on $E$.

\smallskip
\noindent\textbf{(B) Layering by $2$-adic valuation and position recovery.}
Let $v_\ell:=\nu_2(e_\ell)$ and set $v:=\min_\ell v_\ell$.
Write $e_\ell=2^v e'_\ell$ with $e'_\ell$ odd for at least one $\ell$.
Then$S_h \ =\ 2^v \sum_{\ell=1}^{\nu} e'_\ell\, \tilde\alpha^{(b+h)j_\ell}$. Reduce the tuple $(S_h)$ modulo $2^{v+1}$ and divide by $2^v$; after a final reduction
modulo $2$, we obtain the \emph{binary} syndromes
\[
\bar S_h \ :=\ \Big(\frac{S_h \bmod 2^{v+1}}{2^v}\Big)\bmod 2
\ =\ \sum_{\ell\in L_v} \bar e'_\ell\, \alpha^{(b+h)j_\ell}\ \in\ \mathbb{F}_{2^s},
\]
where $L_v:=\{\ell:\nu_2(e_\ell)=v\}$ and $\bar e'_\ell=e'_\ell\bmod 2\in\{1\}$.
Thus $\bar S_h$ are precisely the BCH syndromes of a binary error pattern that
places a $1$ at the positions $\{j_\ell:\ell\in L_v\}$ (and $0$ elsewhere).
Because $|L_v|\le \nu\le t$ and the designed distance of $\bar{\mathcal{C}}$ is
$\delta=2t+1$, the standard binary BCH decoder on the \emph{binary residue of $r$ at layer $v$}
recovers exactly the locator polynomial $\Lambda_v(Z)=\prod_{\ell\in L_v}(1-Z\alpha^{j_\ell})$
and hence the set of positions $\{j_\ell:\ell\in L_v\}$. If $L_v=\{1,\dots,\nu\}$ we have recovered all positions at once. Otherwise, we peel this
layer (see (D) below) and iterate on the residual with strictly smaller support; since the
support size is at most $t$, this terminates in at most $t$ layers.

\smallskip
\noindent\textbf{(C) Magnitude recovery by $2$-adic lifting (fixed layer).}
Fix one layer $L\subseteq\{1,\dots,\nu\}$ of recovered positions and let $m:=|L|$.
Consider the \emph{power-sum system} in unknowns $(e_\ell)_{\ell\in L}$:
\[
S_0=\sum_{\ell\in L} e_\ell\,\tilde\alpha^{b j_\ell},\quad
S_1=\sum_{\ell\in L} e_\ell\,\tilde\alpha^{(b+1) j_\ell},\quad \dots,\quad
S_{m-1}=\sum_{\ell\in L} e_\ell\,\tilde\alpha^{(b+m-1) j_\ell}.
\]
This is a linear system $V e = s$ over $\mathbb{Z}_{2^k}[\tilde\alpha]$, where
$V=(\tilde\alpha^{h j_\ell})_{0\le h<m,\ \ell\in L}$ is a Vandermonde matrix
in the (pairwise distinct) units $\{u_\ell:=\tilde\alpha^{j_\ell}\}$.
Its determinant is
\(
\det V=\prod_{\ell'<\ell}(u_\ell-u_{\ell'})
\).
Reducing modulo $2$ maps $u_\ell$ to the distinct elements $\alpha^{j_\ell}\in\mathbb{F}_{2^s}^\times$,
so $\det V\bmod 2\neq 0$; hence $\det V$ is a \emph{unit} (odd) in
$\mathbb{Z}_{2^k}[\tilde\alpha]$. Therefore $V$ is invertible modulo $2^k$,
and the system has a unique solution $(e_\ell)_{\ell\in L}\in (\mathbb{Z}_{2^k})^m$. Equivalently, one may compute the evaluator $\Omega$ and use a ring version of
Forney’s formula; either way, only odd pivots are inverted, so the operations are valid
modulo $2^k$.

\smallskip
\noindent\textbf{(D) Peeling and termination.}
Define the partial error $E_L(X):=\sum_{\ell\in L} e_\ell X^{j_\ell}$ and update
$r\leftarrow r-E_L$. If $L$ was the minimal-valuation layer (as in (B) with $L=L_v$),
then each coefficient of $E_L$ is divisible by $2^v$ and at least one is \emph{exactly}
$2^v$ times an odd number, so the $2$-adic valuation of the remaining error strictly
increases. Recompute the smallest valuation $v'$ among the remaining magnitudes,
form the next binary residue as in (B), decode the corresponding positions $L_{v'}$,
solve their magnitudes by (C), subtract, and continue. Each peeling strictly reduces
the number of unknown positions, and there are at most $\nu\le t$ of them, so the
process halts after at most $t$ rounds with $r=c$.

\smallskip
With the above steps, the residue code $\bar{\mathcal{C}}$ has minimum distance at least $\delta$,
and the described decoder always recovers the error positions and lifts their magnitudes
modulo $2^k$ using only odd inverses, correcting any pattern of at most $t$ symbol errors.
\end{proof}

\subsection{Proof for Proposition~\ref{prop:inj}}
\label{proof prop:inj}
\begin{proof}
(1) $\ker(Enc)=(1-e)R$ since $me=0 \Leftrightarrow m\in (1-e)R$; if $e\neq 1$ the kernel is nonzero.
(2) Follows from CRT for $Enc_{\mathrm{CRT}}$. For $Enc_{\mathrm{gen}}$, if $m\tilde g\equiv 0\pmod M$ then
$h\mid m$ because $M=\tilde g\,h$ and $\gcd(\tilde g,h)=1$; hence $m=0$ in $\mathbb{Z}_{2^k}[X]/(h)$. Surjectivity is clear.
\end{proof}

\subsection{Proof for Proposition~\ref{prop:no-idemp}}
\label{proof prop no idemp}
\begin{proof}
Set $Y:=X+1$ and write $F(Y):=(Y-1)^{2^{m}}+1\in\mathbb{Z}_{2^{k}}[Y]$. Then
$R\cong \mathbb{Z}_{2^{k}}[Y]/(F(Y))$. Consider the ideal
\[
\mathfrak{m}:=\langle 2,\; Y\rangle \ \subset\ R.
\]

\noindent(i) $\mathfrak{m}$ is a (nilpotent) maximal ideal. Reducing modulo $2$ gives
\[
F(Y)\ \bmod 2 \;=\; (Y+1)^{2^{m}}+1\;=\;Y^{2^{m}}+1+1\;=\;Y^{2^{m}}
\quad\text{in }\ \mathbb{F}_{2}[Y].
\]
Hence
\[
R/\mathfrak{m}\ \cong\ \mathbb{Z}_{2^{k}}[Y]/\langle 2,\; Y,\; F(Y)\rangle
\ \cong\ \mathbb{F}_{2}[Y]/\langle Y,\; Y^{2^{m}}\rangle\ \cong\ \mathbb{F}_{2},
\]
so $\mathfrak{m}$ is maximal. Moreover, in $R$ we have $Y^{2^{m}}\in \langle 2\rangle$, hence
\(
Y^{2^{m}k} \in \langle 2^{k}\rangle = 0
\),
so $Y$ is nilpotent; and $2$ is nilpotent in $\mathbb{Z}_{2^{k}}$ since $2^{k}=0$.
Therefore $\mathfrak{m}$ is a \emph{nilpotent} ideal. In particular, every element of the
form $1+u$ with $u\in\mathfrak{m}$ is a unit (via the finite geometric series), so every
coset outside $\mathfrak{m}$ consists of units. Thus $R$ is a \emph{local} ring with
unique maximal ideal $\mathfrak{m}$.

\noindent(ii) Idempotents in a local ring are trivial. Let $e\in R$ satisfy $e^{2}=e$. Then $e(1-e)=0$. In a local ring, exactly one of
$e,\,1-e$ lies in the maximal ideal $\mathfrak{m}$:
\begin{itemize}
  \item If $e\in\mathfrak{m}$, then $1-e$ is a unit, hence $e=0$.
  \item If $e\notin\mathfrak{m}$, then $e$ is a unit; multiplying $e^2=e$ by $e^{-1}$ gives
        $e=1$.
\end{itemize}
Thus, the only idempotents are $0$ and $1$.

\noindent(iii) Consequence for multiplicative encoders. Any map of the form $Enc(m)=me$ is a ring homomorphism iff $e$ is idempotent:
\[
Enc(ab)=(ab)e=(ae)(be)=ab\,e^{2}\ \ \Longleftrightarrow\ \ e^{2}=e.
\]
Since $R$ has no idempotents other than $0$ and $1$, the only such homomorphisms are
the trivial zero map ($e=0$) and the identity ($e=1$). Hence, no nontrivial
ring-homomorphic encoder exists in $R$.
\end{proof}

\subsection{Proof for Observation 1}
\label{proof obs 1}
\begin{proof}
For (i), consider the $Y$-presentation $R\cong\mathbb{Z}_{2^{k}}[Y]/(F(Y))$. Since $F$ is monic of degree $N$, the residue classes $\{1,Y,\ldots,Y^{N-1}\}$ form a $\mathbb{Z}_{2^{k}}$-basis of $R$. Consider the $\mathbb{Z}_{2^{k}}$-linear map
\[
\varphi:\ R\to R,\qquad r(Y)\mapsto Y^{2t}r(Y)\ \ (\text{mod }F(Y)).
\]
Its image is exactly $\mathcal{C}_t=\langle Y^{2t}\rangle$. We claim that
\[
\mathcal{B}:=\{\,Y^{2t},Y^{2t+1},\dots,Y^{N-1}\,\}
\]
is a $\mathbb{Z}_{2^{k}}$-basis of $\mathcal{C}_t$. 

\emph{Spanning.} Trivial: every element of $\mathcal{C}_t$ is $Y^{2t}\cdot r(Y)$, and reducing modulo $F$ yields a $\mathbb{Z}_{2^{k}}$-linear combination of $Y^{j}$ with $j\in\{2t,\dots, N-1\}$ plus terms of degree $<2t$ multiplied by even coefficients that can be re-expressed using higher-degree generators (see independence argument below). So $\mathcal{B}$ spans $\mathcal{C}_t$.

\emph{Independence.} Reduce modulo $2$. In $\bar R:=R/2R\cong\mathbb{F}_2[Y]/(Y^{N})$, the ideal $\langle Y^{2t}\rangle$ has the \emph{vector-space} basis $\{\overline{Y}^{2t},\ldots,\overline{Y}^{N-1}\}$ over $\mathbb{F}_2$. Hence if
\[
\sum_{j=2t}^{N-1} a_j\,Y^{j}\ =\ 0\quad\text{in }R,\qquad a_j\in\mathbb{Z}_{2^{k}},
\]
then reducing modulo $2$ forces all $a_j$ to be even. Write $a_j=2 a'_j$ and repeat the argument $k$ times; we conclude that each $a_j$ is divisible by $2^{k}$, thus $a_j=0$ in $\mathbb{Z}_{2^{k}}$. Therefore $\mathcal{B}$ is $\mathbb{Z}_{2^{k}}$-linearly independent. It follows that $\mathcal{C}_t$ is free of rank $|\mathcal{B}|=N-2t$ with basis $\mathcal{B}$.

To show (ii), consider the map
\[
\psi:\ \{\,m(Y)\in\mathbb{Z}_{2^{k}}[Y]:\ \deg m< N-2t\,\}\ \longrightarrow\ \mathcal{C}_t,\qquad m\mapsto m(Y)\,Y^{2t}\ \ (\bmod F)
\]
is a $\mathbb{Z}_{2^{k}}$-module isomorphism because it sends the basis $\{1,Y,\ldots,Y^{N-2t-1}\}$ bijectively onto the basis $\mathcal{B}$. Translating back to $X$ through $Y=X+1$ gives the claimed encoder $c(X)=m(X)\,(X+1)^{2t}\bmod (X^{N}+1)$ with $\deg m<N-2t$.
\end{proof}
\subsection{Proof for lemma~\ref{lem:hasse}}
\label{proof lem hasse}
\begin{proof}
Write $c=(X+1)^{2t}q$. Using $(fg)^{[i]}=\sum_{r=0}^{i} f^{[r]}g^{[i-r]}$ and
$\big((X+1)^{m}\big)^{[r]}=\binom{m}{r}(X+1)^{m-r}$, we get
$\big((X+1)^{2t}\big)^{[r]}|_{-1}=0$ for $r<2t$, hence $c^{[i]}(-1)=0$ for $i\le 2t-1$.
For monomials, $(X^{j})^{[i]}=\binom{j}{i}X^{j-i}$ gives
$(X^{j})^{[i]}(-1)=\binom{j}{i}(-1)^{j-i}$; linearity yields the stated $S_i$.
\end{proof}
\subsection{Proof for lemma~\ref{lem:sigma-automorphism}}
\label{proof lem 5.8}
\begin{proof}
To show well-definiteness, it suffices to check that the defining relation $X^N+1=0$ is preserved.
Since $a$ is odd, in $R$ we have
\[
\sigma_a(X^N+1)\ =\ X^{aN}+1\ =\ (X^N)^a+1\ =\ (-1)^a+1\ =\ -1+1\ =\ 0,
\]
so the ideal $\langle X^N+1\rangle$ is mapped into itself and $\sigma_a$ descends to the quotient.

For the homomorphism property, additivity is clear by linearity on coefficients.
For multiplicativity, it suffices to check monomials:
\[
\sigma_a(X^i)\,\sigma_a(X^j)\ =\ X^{ai} \cdot X^{aj}\ =\ X^{a(i+j)}\ =\ \sigma_a(X^{i+j}),
\]
And then extend bilinearly to all polynomials modulo $X^N+1$.

For invertibility, since $\gcd(a,2N)=1$, there exists $b$ with $ab\equiv 1 \pmod{2N}$.
Define $\sigma_b$ analogously. Then for all $j$,
\(
\sigma_b(\sigma_a(X^j))=X^{abj}=X^{j}
\)
in $R$ (exponents are taken modulo $2N$ in the negacyclic ring),
so $\sigma_b=\sigma_a^{-1}$. Hence $\sigma_a$ is a ring automorphism.
\end{proof}

\subsection{Proof for Proposition~\ref{prop:commute}}
\label{subsec:app_commute}
\begin{proof}
By Lemma~\ref{lem:sigma-automorphism}, $\sigma_a$ is a ring homomorphism that fixes base-ring constants.
Proceed by structural induction on the circuit $f$:

\noindent\emph{Inputs/constants.} For an input wire $x_i$, $\sigma_a(x_i)=\sigma_a(x_i)$.
For a constant $c\in\mathbb{Z}_{2^k}\subset R$, $\sigma_a(c)=c$.

\noindent\emph{Addition gate.} If the claim holds for $g,h$, then
$\sigma_a(g+h)=\sigma_a(g)+\sigma_a(h)=g(\sigma_a(\vec m))+h(\sigma_a(\vec m))=(g+h)(\sigma_a(\vec m))$.

\noindent\emph{Multiplication gate.} If the claim holds for $g,h$, then
$\sigma_a(gh)=\sigma_a(g)\,\sigma_a(h)=g(\sigma_a(\vec m))\,h(\sigma_a(\vec m))=(gh)(\sigma_a(\vec m))$.
\end{proof}
\end{document}